\documentclass[prd,preprintnumbers,showpacs,onecolumn,showkeys,floatfix,nofootinbib,notitlepage]{revtex4-1}
\usepackage{graphicx}
\usepackage{amssymb}
\usepackage{amsmath}
\usepackage{mathtools}
\usepackage{epsfig}
\usepackage{color}
\usepackage{enumerate}
\usepackage{slashed}
\usepackage{hyperref}
\usepackage{epstopdf}

\def\slashchar#1{\setbox0=\hbox{$#1$}
   \dimen0=\wd0 \setbox1=\hbox{/} \dimen1=\wd1
   \ifdim\dimen0\big>\dimen1 \rlap{\hbox to \dimen0{\hfil/\hfil}} #1
   \else  \rlap{\hbox to \dimen1{\hfil$#1$\hfil}} / \fi}

\newcommand{\ud}{\mathrm{d}}
\newcommand{\be}{\begin{equation}}
\newcommand{\ee}{\end{equation}}
\newcommand{\bea}{\begin{eqnarray}}
\newcommand{\eea}{\end{eqnarray}}

\newcommand{\Appendix}[1]%
    {%
     \section{#1}%
      }

\begin{document}

\title{Cancellation of Infrared Divergence in Inclusive Production of Heavy Quarkonia}

\author{Gao-Liang Zhou
}
\affiliation{Center for High Energy Physics, Peking University,  Beijing 100190, People's Republic of China}




\begin{abstract}
A scheme is presented here to cancel out topologically unfactorized infrared divergences in the inclusive production of heavy quarkonia, which affect the nonrelativistic QCD (NRQCD) factorization of these processes. Heavy quarkonia are defined as resonance states of QCD instead of color singlet heavy quark pair. Thus the final heavy quark pair is not necessarily a color singlet. In addition, Heavy quarkonia are reconstructed from their decay products. As a result, transition between states containing heavy quarks caused by exchanges of soft gluons are also taken into account here.   Such cancellation is crucial for the NRQCD factorization of these processes.
\end{abstract}

\maketitle

\section{Introduction}
\label{introduction}

The production of heavy quarkonia forms an important and interesting issue in the study of QCD and the strong interaction\cite{Brambilla:2010cs,BBEOP2013}. The large mass of heavy quarks suggests that one may treat heavy quarkonia as non-relativistic bound systems. This is supported by quark potential model calculations, which indicates that $v^{2}\sim 0.3$ for charmonium and $v^{2}\sim 0.1$ for bottomonium\cite{Quigg:1979vr} with $v$ the typical velocity in the center- of-mass frame of heavy quarkonia. It has been proposed that the effective theory non-relativistic QCD(NRQCD)\cite{Caswell:1985ui,Bodwin:1994jh} could be used to describe the separation of short- and long-distance effects in the production of heavy quarkonia. Short-distance effects that produce a heavy quark pair are perturbative, while long distance effects that evolute the heavy quark pair into heavy quarkonium are non-perturbative and independent of explicit process. The factorization theorem proposed in \cite{Bodwin:1994jh} can be written as
\begin{equation}
\ud \sigma_{A+B\to H+X}=\sum_{n}\ud \sigma_{ A+B\to Q\bar{Q}(n)+X}\big<\mathcal{O}^{H}(n)\big>
\end{equation}
, where $A$ and $B$ represent initial particles, $H$ represents the detected heavy quarkonium, $X$ represents undetected final particles, $Q\bar{Q}(n)$ represents the heavy quark pair in a special color and angular momentum state.

Although the NRQCD factorization formula is widely used in the study of production of heavy quarkonia and has gained great success in explanation of experimental data(see, for example, Refs.
\cite{Braaten:1994vv,Cho:1995vh,Cho:1995ce,Amundson:1996ik,Beneke:1996yw,Braaten:1996jt,Cacciari:1996dg,Huang:1996cs,Braaten:1999qk,Braaten:2000cm,
Gong:2007db,Ma:2010yw,Butenschoen:2011yh,Gong:2013qka}), there remain challenges that NRQCD factorization formula is facing, particularly the $J/\psi$ puzzle and the surprisingly large cross section of the associated production of $J/\psi$ in $e^{+}e^{-}$(see, for example, Ref.\cite{Brambilla:2010cs} and references there in). It seems that more efforts are needed to examine the NRQCD factorization formula, especially for inclusive production of heavy quarkonia.

Proofs of NRQCD factorization theorem for the exclusive production of two charmonia in $e^{+}e^{-}$ annihilation and production of a charmonium and a light meson in B-meson decays are established in \cite{Bodwin:2008nf,Bodwin:2010fi}. For inclusive production of heavy quarkonia, however, the issue is quite nontrivial\cite{NQS:2005,Nayak:2006fm}. In \cite{Bodwin:1994jh}, it was argued that infrared divergence caused by exchanges of soft gluons between heavy quarks and extra jets cancel out once the summation  over undetected particles has been made, even while one constrain the final heavy quark pair to be color singlet. According to explicit calculations at next-to-next-to-leading order(NNLO), authors in \cite{NQS:2005,Nayak:2006fm} find that it is necessary to modify NRQCD octet production matrix elements to include non-Abelian phases(which makes them gauge invariant)in order to restore NRQCD factorization in inclusive production of heavy quarkonia.

In spite of difficulties in the proof of NRQCD factorization theorem for inclusive production of heavy quarkonia, it is proved\cite{Kang:2011mg,Kang:2011zza,Kang:2014tta,Fleming:2012wy,Fleming:2013qu}that collinear factorization holds up to order $M^{2}/p_{T}^{2}$ for  such processes with $M$ the mass of heavy quark and $p_{T}$ the transverse momenta of the detected heavy quarkonia in the mass center frame of initial particles. In the collinear factorization approach, the differential cross section of inclusive production of heavy quarkonium reads
\begin{eqnarray}
\label{collinear factorization}
\ud \sigma_{A+B\to H+X}&=&\sum_{i}\ud \sigma_{A+B\to i+X}\otimes D_{i\to H} +
\sum_{\kappa} \ud \sigma_{ A+B\to Q\bar{Q}(\kappa)+X}\otimes D_{Q\bar{Q}(\kappa)\to H}
\nonumber\\
&&+\mathcal{O}(M^{4}/p_{T}^{4})
\end{eqnarray}
, where $\ud \sigma_{A+B\to i+X}$ represents the differential cross section of inclusive production of a parton $i$, $D_{i\to H}$ represents the fragmentation function of $i$ into $H$ with $i$ produced in a short distance process(order $1/p_{T}$), $\ud \sigma_{ A+B\to Q\bar{Q}(\kappa)+X}$
represents the differential cross section of inclusive production of a heavy quark pair $Q\bar{Q}$ in a special color and angular momentum state $\kappa$, $D_{Q\bar{Q}(\kappa)\to H}$ represents the fragmentation function of  the heavy quark pair $Q\bar{Q}$ into $H$ with the pair produced in a short distance process(order $1/p_{T}$). The first term in \ref{collinear factorization} contributes from leading power in $M/p_{T}$, while the second term  in \ref{collinear factorization} contributes from subleading power in $M/p_{T}$. If NRQCD factorization holds up to order $M^{2}/p_{T}^{2}$, then one has\cite{Kang:2011mg,Kang:2014tta}
\begin{eqnarray}
\label{fragmentation}
D_{i\to H}&=& \sum_{n}d_{i\to Q\bar{Q}(n)}\big<\mathcal{O}^{H}(n)\big>
\nonumber\\
D_{Q\bar{Q}(\kappa)\to H}&=&\sum_{n} d_{Q\bar{Q}(\kappa)\to Q\bar{Q}(n)}\big<\mathcal{O}^{H}(n)\big>
\end{eqnarray}
, where $d_{i\to Q\bar{Q}(n)}$ and $d_{Q\bar{Q}(\kappa)\to Q\bar{Q}(n)}$ are short distance(order $1/M$) coefficients.

\begin{figure}
\begin{center}
\includegraphics[width=0.4\textwidth]{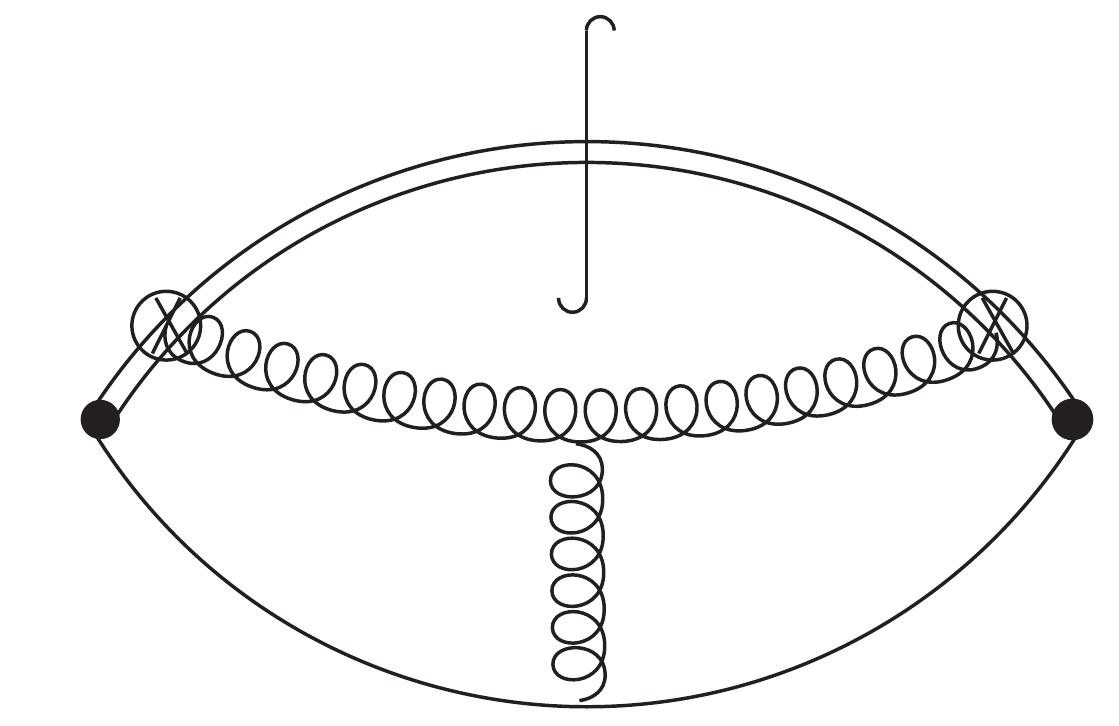}
\end{center}
\caption{Topology of diagrams which contribute to uncancelled infrared divergence that affect NRQCD factorization at NNLO. Summation over all cuts that produce a color singlet heavy quark pair, which is produced as color octet at the hard vertex, is understood implicitly. }
\label{fig:infradiv}
\end{figure}

In this paper, we present a scheme to cancel out topologically unfactorized infrared divergences exhibited in \cite{NQS:2005,Nayak:2006fm}.  In \cite{NQS:2005,Nayak:2006fm}, it was shown that the uncancelled infrared divergence that affect NRQCD factorization  originates from diagrams of the type that shown in Fig.\ref{fig:infradiv}.
For an Abelian gauge theory like QED, diagrams of this type do not appear. Thus factorization  is not bothered by these diagrams for Abelian gauge theory. One may ask, what happens for non-Abelian gauge theory? For non-Abelian gauge theory, the color of the heavy quark pair is affected by soft gluons which couple to the pair. Once the color state of the final heavy quark pair is fixed, we can no longer make the  the inclusive summation over all states made up of a heavy quark pair and an arbitrary number of infrared gluons. In fact,  the final heavy quark pair is constrained to be color singlet in diagrams of the type shown in Fig.\ref{fig:infradiv}. Thus the KLN cancellation does not simply work in this case. It is necessary to point it out, however, that the color state of final heavy quark pair is not fixed in actual process. What we detected are heavy quarkonia like $J/\psi$ not color singlet heavy quark pair in actual process. Infrared behaviours of higher Fock stats should also be taken into account. Summation over these Fock states, however, is not enough to cancel out topologically unfactorized infrared divergences as we will see according to explicit calculations. Exchanges of soft gluons between the heavy quarks and undetected states $X$ may cause  transition between states containing heavy quarks even though momenta of soft gluons tend to 0! Fortunately, heavy quarkonia are reconstructed from their decay products in actual experiments. As a result, we consider a more inclusive process:
\begin{equation}
A+B\to\mu^{+}\mu^{-}(n,p_{H})+X
\end{equation}
, where $\mu^{+}\mu^{-}(n,p_{H})$ is a $\mu^{+}\mu^{-}$ pair with the intrinsic quantum numbers equal to those of the detected heavy quarkonium $H$. The invariant mass of the $\mu^{+}\mu^{-}$ pair is constrained to be near the mass of the heavy quarkonium $H$. We will show that
topologically unfactorized infrared divergences do cancel out in such a process.

The paper is organized as  follows. In Sec.\ref{incp}, we describe the inclusive production process of heavy quarkonia.  In Sec.\ref{Jpsi}, we take the $J/\psi$ particle as an example to explain how to define Heavy quarkonia so that they are invariant under the evolution of infrared QCD interactions.  In Sec.\ref{cancinf}, we show that summation over higher Fock states is not enough to cancel out topologically unfactorized infrared divergences according to explicit calculations at NNLO. In Sec.\ref{incpro}, we consider the inclusive production of $\mu^{+}\mu^{-}$ pair with invariant mass near the mass of a heavy quarkonium $H$ and intrinsic quantum numbers equal to those of $H$. We show that    topologically unfactorized infrared divergences do cancel out in this process.     We give our conclusion and some discussions in Sec.\ref{conclusion}.

\section{Inclusive Production of Heavy Quarkonia }
\label{incp}

In this section, we describe the inclusive production process of heavy quarkonia. The process can be written as
\begin{equation}
A(p_{A})+B(p_{B})\to H(p_{H})+X(p_{X})
\end{equation}
, where $A$ and $B$ represent initial particles, $X$ represents undetected final particles.
In the center- of-mass frame of the initial particles, momenta of $A$ and $B$ are nearly light like. We have
\begin{equation}
p_{A}^{\mu}=\bar{n}\cdot p_{A}n^{\mu}+n\cdot p_{A}\bar{n}^{\mu}
          \simeq \bar{n}\cdot p_{A}n^{\mu}
\end{equation}
\begin{equation}
p_{B}^{\mu}=n\cdot p_{B}\bar{n}^{\mu}+\bar{n}\cdot p_{B}n^{\mu}
\simeq n\cdot p_{B}\bar{n}^{\mu}
\end{equation}
, where $n^{\mu}$ and $\bar{n}^{\mu}$ are light-like vectors:
\begin{equation}
n^{\mu}=\frac{1}{\sqrt{2}}(1,\vec{n}),\quad \bar{n}^{\mu}=\frac{1}{\sqrt{2}}(1,-\vec{n})
\end{equation}
. Transverse momentum of the final heavy quarkonium $H$ is proposed to be much greater than the mass of $H$. In this case, the momentum of $H$ is also nearly light-like in the  center- of-mass frame of initial particles. We have:
\begin{eqnarray}
p_{H}^{\mu}&\equiv& \bar{n_{H}}\cdot p_{H}n_{H}^{\mu}+n_{H}\cdot p_{A}\bar{n_{H}}^{\mu}
 +(p_{H}^{\mu}-\bar{n_{H}}\cdot p_{H}n_{H}^{\mu}+n_{H}\cdot p_{A}\bar{n_{H}}^{\mu})
 \nonumber\\
          &\simeq& \bar{n_{H}}\cdot p_{H}n_{H}^{\mu}
\end{eqnarray}
, where $n_{H}^{\mu}$ and $\bar{n_{H}}^{\mu}$ are light like vectors with:
\begin{equation}
n_{H}^{\mu}=\frac{1}{\sqrt{2}}(1,\vec{n_{H}}),\quad \bar{n_{H}}^{\mu}=\frac{1}{\sqrt{2}}(1,-\vec{n_{H}}),\quad
\end{equation}
. We propose that $n_{H}\cdot n\sim n_{H}\cdot \bar{n}\sim 1$ in this paper.

The collinear factorization theorem for the process reads(\cite{Kang:2011mg,Kang:2011zza,Kang:2014tta,Fleming:2012wy,Fleming:2013qu})
\begin{eqnarray}
\ud \sigma_{A+B\to H+X}&=&\sum_{i}\ud \sigma_{A+B\to i+X}\otimes D_{i\to H} +
\sum_{\kappa} \ud \sigma_{ A+B\to c\bar{c}(\kappa)+X}\otimes D_{c\bar{c}(\kappa)\to H}
\nonumber\\
&&+\mathcal{O}(M^{4}/p_{T}^{4})
\end{eqnarray}
, where fragmentation functions $D_{i\to H}$ and $D_{c\bar{c}(\kappa)\to H}$ are defined in terms of expectation values of non-local operators between vacuumed and final hadron states.  For example, the bare fragmentation function of light quark reads(\cite{CSS:1982pd}):
\begin{eqnarray}
D_{q\to H}^{(0)}(z)
&=&\frac{z^{d-3}}{24\pi}
\int\ud n_{H}\cdot y\quad e^{-i\bar{n_{H}}\cdot p_{H} n_{H}\cdot y/z}
\nonumber\\
&&
\sum_{Y}Tr \big<0|
       W_{n_{H}}^{\dag}\psi(0,n_{H}\cdot y,\vec{0})
|H Y\big>
 \big<H Y
   |\psi^{\dag}W_{n_{H}}(0)
   |0\big>
\end{eqnarray}
, where $z$ is the ratio of the large momentum component of $H$ to that of the light quark $i$, $W_{n_{H}}$ is the light like Wilson line:
\begin{equation}
W_{n_{H}}(x)=(\mathcal{P}\exp(ig\int_{0}^{\infty}\ud s
\bar{n_{H}}\cdot A(x+s\bar{n_{H}}))^{\dag}
\end{equation}

If NRQCD factorization holds up to order $M^{2}/p_{T}^{2}$, then one has\cite{Kang:2011mg,Kang:2014tta}:
\begin{eqnarray}
D_{i\to H}&=& \sum_{n}d_{i\to Q\bar{Q}(n)}\big<\mathcal{O}^{H}(n)\big>
\nonumber\\
D_{Q\bar{Q}(\kappa)\to H}&=&\sum_{n} d_{Q\bar{Q}(\kappa)\to Q\bar{Q}(n)}\big<\mathcal{O}^{H}(n)\big>
\end{eqnarray}
, where $d_{i\to Q\bar{Q}(n)}$ and $d_{Q\bar{Q}(\kappa)\to Q\bar{Q}(n)}$ are short distance coefficients. $d_{i\to Q\bar{Q}(n)}$ and $d_{Q\bar{Q}(\kappa)\to Q\bar{Q}(n)}$ should be infrared safe, that is to say, all infrared divergences in $D_{i\to H}$ and $D_{Q\bar{Q}(\kappa)\to H}$ should be absorbed in to the long distance matrix elements of the effective operators $\mathcal{O}^{H}(n)$.  Calculations in \cite{NQS:2005,Nayak:2006fm} show that, however, the infrared safety of $d_{i\to Q\bar{Q}(n)}$ and $d_{Q\bar{Q}(\kappa)\to Q\bar{Q}(n)}$ is much non-trivial once the final detected particle is the color singlet heavy quark pair.

It is necessary to point out that the final detected particle is $H$ not color singlet heavy quark pair in the process considered here. The color-singlet heavy quark pair is not invariant under the evolution of QCD even though annihilation of  heavy quarks are neglected. The $H$ particle, however, is a stable particle once the decay of $H$  is neglected. Thus effects of higher Fock states are important in the evolution of heavy quarkonia under infrared QCD interactions.

\section{Relations Between $J/\psi$ and Heavy Quark Pair}
\label{Jpsi}

In this section, we take the $J/\psi$ particle as an example to show how to define heavy qaurkonia so that they are invariant under infrared QCD interactions once decay of heavy quarknia is neglected.
$J/\psi$ is a stable particle if one neglect the annihilation of the charm quark pair. We thus define the $J/\psi$ state as eigenstate of NRQCD, where effects of electroweak interactions on the structure of $J/\psi$ are neglected here. We have:
\begin{eqnarray}
\vec{P}|J/\psi(j_{z},\vec{p})\big>&=&\vec{p}|J/\psi(j_{z},\vec{p})\big>
\\
J^{2}|J/\psi(j_{z},\vec{0})\big>&=&2|J/\psi(j_{z},\vec{0})\big>
\\
J_{z}|J/\psi(j_{z},\vec{0})\big>&=&j_{z}|J/\psi(j_{z},\vec{0})\big>
\\
\label{eigeneq}
H_{NRQCD}^{(h)}|J/\psi(j_{z},\vec{0})\big>&=& (M_{J/\psi}-2M_{c})|J/\psi(j_{z},\vec{0})\big>
\end{eqnarray}
, where $\vec{p}$ represents the momentum of the $J/\psi$ state, $j_{z}$ represents the $z$ component of the spin of the $J/\psi$ state, $\vec{P}$ and $\vec{J}$ represents the momentum operator and the angular momentum operator respectively, $M_{J/\psi}$ and $M_{c}$ represent the mass of $J/\psi$ and charm quark respectively, $H_{NRQCD}^{(h)}$ represents the hermitian part of the Hamiltonian of NRQCD. We do not consider the anti-hermitian part, which describes effects of annihilation of the heavy quark pair, as the width of $J/\psi$ is much smaller than the binding energy of $J/\psi$.

We expand the $J/\psi$ state according to eigenstates of $H_{NRQCD}^{(0)}$ that is the free part of $H_{NRQCD}$ and have:
\begin{equation}
|J/\psi(j_{z},\vec{0})\big>=\sum_{m_{1},m_{2}}\int\frac{\ud^{3}q}{(2\pi)^{3}}
\phi(\vec{q},j_{z},m_{1},m_{2})|c(\vec{q},m_{1})\bar{c}(-\vec{q},m_{2})\big>+\text{high Fock states}
\end{equation}
, where  $|c(\vec{q},m_{1})\bar{c}(-\vec{q},m_{2})\big>$ is a color singlet charm pair, $m_{i}=\frac{1}{2}, -\frac{1}{2}$($i=1,2$), $\phi(\vec{q},j_{z},m_{1},m_{2})$ is the wave function:
\begin{equation}
\label{wavefunction}
\phi(\vec{q},j_{z},m_{1},m_{2})=\big<c(\vec{q},m_{1})\bar{c}(-\vec{q},m_{2})|J/\psi(j_{z},\vec{0})\big>
\end{equation}
. We bring in the notations
\begin{equation}
\label{ccbarstate}
|c\bar{c},j_{z}\big>_{J/\psi}\equiv \sum_{m_{1},m_{2}}\int\frac{\ud^{3}q}{(2\pi)^{3}}
\phi(\vec{q},j_{z},m_{1},m_{2})|c(\vec{q},m_{1})\bar{c}(-\vec{q},m_{2})\big>
\end{equation}
\begin{equation}
|h,j_{z}\big>\equiv |J/\psi(j_{z},\vec{0})\big>-|c\bar{c},j_{z}\big>_{J/\psi}
\end{equation}
and have 
\begin{equation}
|J/\psi(j_{z},\vec{0})\big>=|c\bar{c},j_{z}\big>_{J/\psi}+|h,j_{z}\big>
\end{equation}
\begin{equation}
\big<h,j_{z}|c\bar{c},j_{z}\big>_{J/\psi}=0
\end{equation}
.  The wave function $\phi(\vec{q},j_{z},m_{1},m_{2})$ is nonperturbative, which is treated as an undetermined function in this paper. We assume that the lowest Fock state is the dominate part in constitutes of $J/\psi$ as people do in \cite{Bodwin:1994jh}. In terms of perturbation theory, this is equivalent to define $J/\psi$ as bound state near the lowest Fock state.

We consider the evolution of the state $|c\bar{c},j_{z}\big>_{J/\psi}$:
\begin{eqnarray}
e^{-iH_{NRQCD}^{(h)}t}|c\bar{c},j_{z}\big>_{J/\psi}
&=&e^{-i(M_{J/\psi}-2M_{c})t} \big<J/\psi(j_{z},\vec{0})|c\bar{c},j_{z}\big>_{J/\psi}
|J/\psi(j_{z},\vec{0})\big>
\nonumber\\
&&
+\sum_{n=2}^{\infty}e^{-iE_{n}t}\big<n,j_{z},\vec{0}|c\bar{c},j_{z}\big>_{J/\psi}
|n,j_{z},\vec{0}\big>
\end{eqnarray}
, where $|n\big>$ represents the spin one eigenstate of $H_{NRQCD}^{(h)}$ that contain a heavy quark pair. $J/\psi$ is the state with smallest invariant mass among these states, we thus have:
\begin{equation}
E_{n}>M_{J/\psi}-2M_{c}\quad (n\ge 2)
\end{equation}
.  We then have:
\begin{eqnarray}
\lim_{t\to \infty(1-i\epsilon)}e^{-i(H_{NRQCD}^{(h)}-M_{J/\psi}+2M_{c})t}|c\bar{c},j_{z}\big>_{J/\psi}
&=&\big<J/\psi(j_{z},\vec{0})|c\bar{c},j_{z}\big>_{J/\psi}
|J/\psi(j_{z},\vec{0})\big>
\end{eqnarray}
We thus define the $J/\psi$ state as:
\begin{equation}
\label{jpsistate}
|J/\psi(j_{z},\vec{0})\big>=\frac{1}{\big<J/\psi(j_{z},\vec{0})|c\bar{c},j_{z}\big>_{J/\psi}}
\lim_{t\to \infty(1-i\epsilon)}e^{-i(H_{NRQCD}^{(h)}-M_{J/\psi}+2M_{c})t}|c\bar{c},j_{z}\big>_{J/\psi}
\end{equation}
. According to (\ref{ccbarstate}) and (\ref{wavefunction}), we have:
\begin{eqnarray}
\big<J/\psi(j_{z},\vec{0})|c\bar{c},j_{z}\big>_{J/\psi}&=&\sum_{m_{1},m_{2}}\int\frac{\ud^{3}q}{(2\pi)^{3}}
|^{2}\phi(\vec{q},j_{z},m_{1},m_{2})|
\nonumber\\
&=&\sum_{m_{1},m_{2}}\int\frac{\ud^{3}q}{(2\pi)^{3}}
|\big<c(\vec{q},m_{1})\bar{c}(-\vec{q},m_{2})|J/\psi(j_{z},\vec{0})\big>|^{2}
\end{eqnarray}
. The undetermined parameter $\big<J/\psi(j_{z},\vec{0})|c\bar{c},j_{z}\big>_{J/\psi}$ is independent of the Fock states in the perturbation series. It can be dropped in the following calculations.

According to (\ref{ccbarstate}) and (\ref{jpsistate}), we have:
\begin{eqnarray}
\label{jpsi-ccbar}
|J/\psi(j_{z},\vec{0})\big>&=&\frac{1}{\big<J/\psi(j_{z},\vec{0})|c\bar{c},j_{z}\big>_{J/\psi}}
\sum_{m_{1},m_{2}}\int\frac{\ud^{3}q}{(2\pi)^{3}}\phi(\vec{q},j_{z},m_{1},m_{2})
\nonumber\\
&&
\lim_{t\to \infty(1-i\epsilon)}
e^{-iH_{NRQCD}^{(0)}t}e^{iH_{NRQCD}^{(0)}t}
\nonumber\\
&&
e^{-i(H_{NRQCD}^{(h)}-M_{J/\psi}+2M_{c})t}
|c(\vec{q},m_{1})\bar{c}(-\vec{q},m_{2})\big>
\nonumber\\
&=&
\frac{1}{\big<J/\psi(j_{z},\vec{0})|c\bar{c},j_{z}\big>_{J/\psi}}
\sum_{m_{1},m_{2}}\int\frac{\ud^{3}q}{(2\pi)^{3}}\phi(\vec{q},j_{z},m_{1},m_{2})
\nonumber\\
&&
\lim_{t\to \infty(1-i\epsilon)}
e^{-i(H_{NRQCD}^{(0)}-M_{J/\psi}+2M_{c})t}
\nonumber\\
&&
(T\{\exp(-i\int_{0}^{t}\ud t^{\prime}H_{NRQCD}^{(I)}(t^{\prime}))\})
|c(\vec{q},m_{1})\bar{c}(-\vec{q},m_{2})\big>
\end{eqnarray}
, where $H_{NRQCD}^{(0)}$  is the free part $H_{NRQCD}$, $E_{q}$ is the energy of the charm quark with momentum $\vec{q}$:
\begin{equation}
E_{q}=\sqrt{M_{c}^{2}+|\vec{q}|^{2}}
\end{equation}
 $H_{NRQCD}^{(I)}(t^{\prime})$ is defined as:
\begin{equation}
H_{NRQCD}^{(I)}(t^{\prime})=e^{iH_{NRQCD}^{(0)}t^{\prime}}(H_{NRQCD}^{(h)}-H_{NRQCD}^{(0)})e^{-iH_{NRQCD}^{(0)}t^{\prime}}
\end{equation}
. We notice that the $i\epsilon$ term in (\ref{jpsi-ccbar}) is in accordance with the Feynman boundary conditions. We can thus calculate the perturbation series of (\ref{jpsi-ccbar}) according to Feyman diagram skills.

\section{Contributions of Higher Fock States}
\label{cancinf}

In this section, we consider contributions of higher Fock state of the $J/\psi$ particle in fragmentation functions $D_{i\to J/\psi}$ and $D_{c\bar{c}(\kappa)\to J/\psi}$ up to NNLO.

The term  $e^{-i(H_{NRQCD}^{(0)}-M_{J/\psi}+2M_{c})t}$
is independent of Fock states once the relative momenta of the final heavy quark pair is definite in the infrared limit. We thus simply drop such term in following calculations.
To determine infrared divergences caused by  soft gluons exchanged between $J/\psi$ and other energetic particles, we take the eikonal line approximation in couplings of soft gluons to the on-shell charm quaik pair. This is equivalent to absorb effects of these soft gluons into Wilson lines:
\begin{equation}
Y_{v}(0)=\mathcal{P}\exp(-ig \int_{0}^{\infty}\ud s v\cdot A(sv))
\end{equation}
, where
\begin{equation}
v^{\mu}=\frac{p^{\mu}}{2p^{0}}
\end{equation}
with $p^{\mu}$ momenta of final heavy quarks.

The gauge invariant effective operators defined in \cite{NQS:2005,Nayak:2006fm} read
\begin{equation}
\mathcal{O}=\sum_{X,j_{z}}\chi^{\dag}Y_{l}^{\dag}(0)\mathcal{K} Y_{l}(0)\psi(0)|J/\psi(j_{z},\vec{0})X\big>\big<J/\psi(j_{z},\vec{0})X\big|\psi^{\dag}Y_{l}^{\dag}(0)\mathcal{K}^{\prime}Y_{l}(0)\chi(0),
\end{equation}
, where $\psi$ is the Pauli spinor field that annihilate the charm quark, $\chi$ is the Pauli spinor field that create the anticharm quark, $\mathcal{K}$ and $\mathcal{K}^{\prime}$ are possible color, spin and covariant derivative terms.  We consider the evolution of a color octet charm pair to the $J/\psi$ state. The infrared behaviour of such evolution can be written as:
\begin{equation}
\label{matrix8-1}
\mathcal{O}^{(8)}(v_{1},v_{2};v_{1}^{\prime}, v_{2}^{\prime})=\sum_{X}\big<0|(Y_{v_{2}}^{\dag}t^{a}Y_{v_{1}})_{ij}Y_{l}^{\dag}(0)_{ac}|J/\psi X\big>
\big<J/\psi X|Y_{l}(0)_{cb}(Y_{v_{2}^{\prime}}^{\dag}t^{b}Y_{v_{1}^{\prime}})^{\dag}_{kl}|0\big>
\end{equation}
, where $v_{1}$ and $v_{1}^{\prime}$ are velocities of final charm quarks, $v_{2}$ and $v_{2}^{\prime}$ are velocities of final anticharm quarks. $v_{i}\ne v_{i}^{\prime}$($i=1,2$) in general case. We do not require that $i=j$ or $k=l$ as the final charm quark pair can be a color octet.

\begin{figure*}
\begin{tabular}{c@{\hspace*{10mm}}c@{\hspace*{10mm}}c}
\includegraphics[scale=0.3]{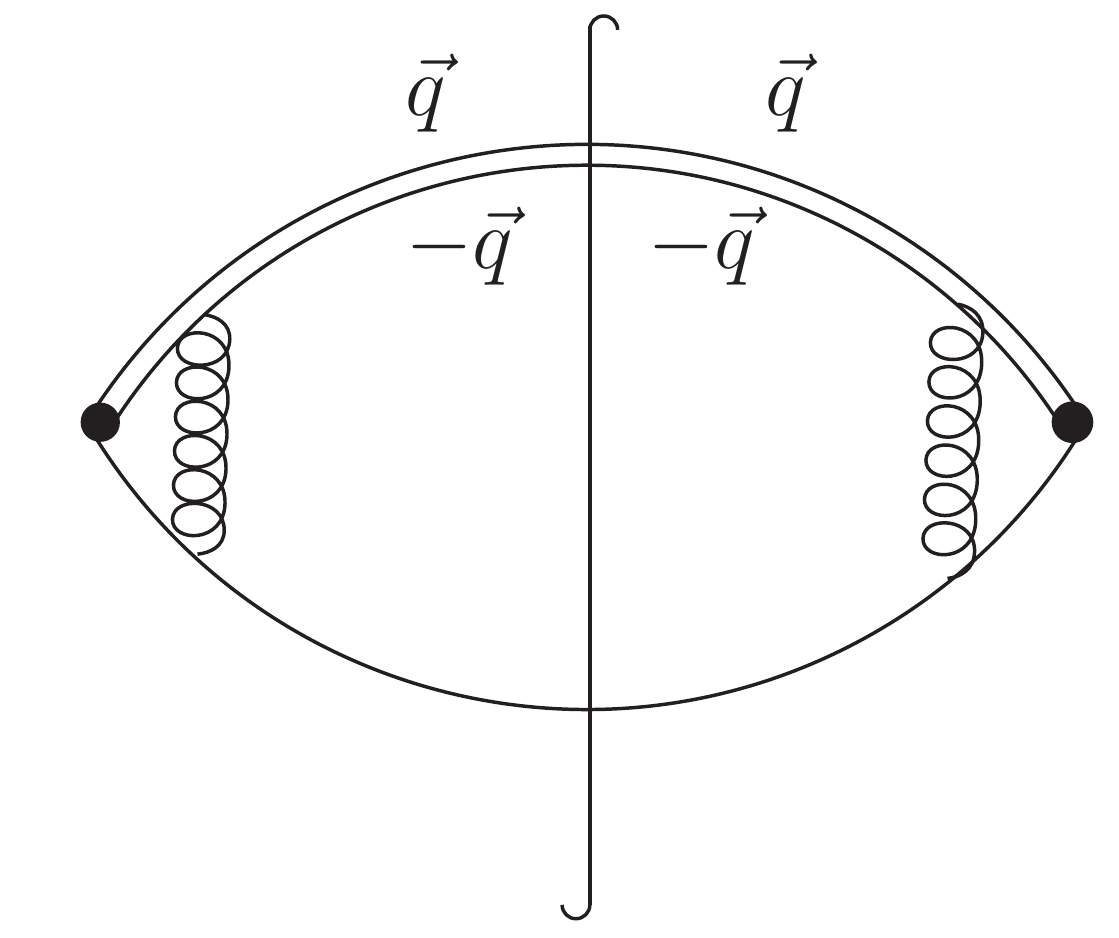}
&
\includegraphics[scale=0.3]{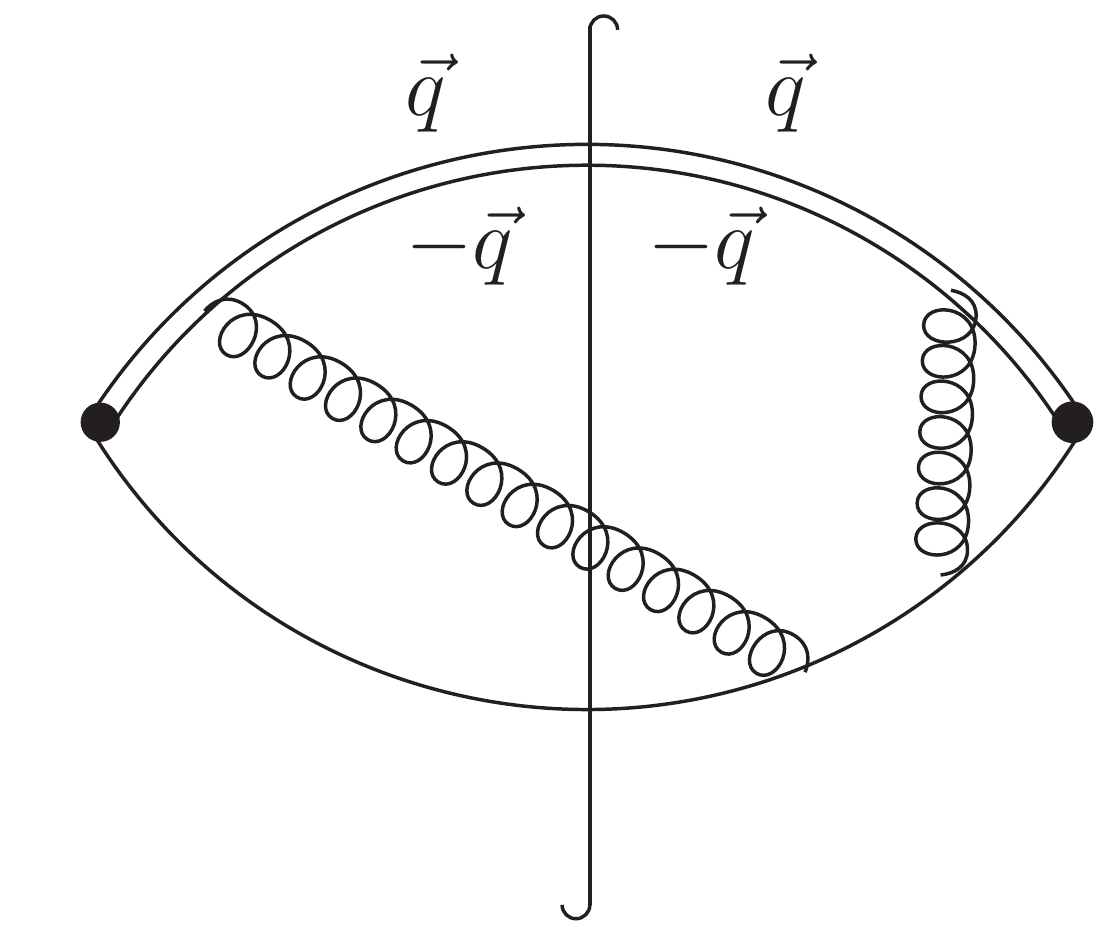}
&
\includegraphics[scale=0.3]{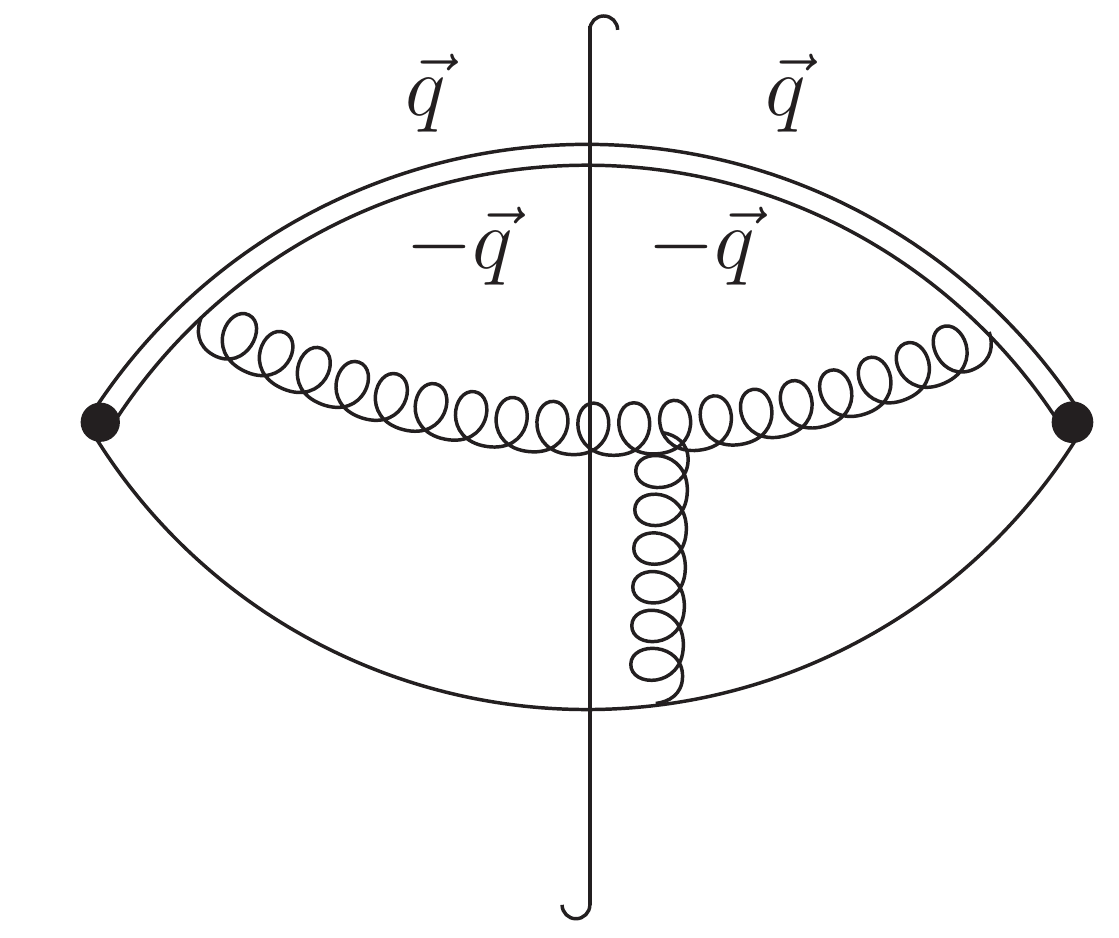}
\\
(a)&(b)&(c)
\end{tabular}
\caption{Examples of diagrams that a color singlet $c\bar{c}$ pair is produced by the evolution of a color octet $c\bar{c}$ pair.}
\label{1fock}
\end{figure*}

Examples of diagrams-of which the lowest Fock states in (\ref{jpsi-ccbar})are produced by the evolution of the color octet heavy quark pair-are shown in Fig.\ref{1fock}.
In fig.\ref{1fock}, we take the same value for the relative momenta of the charm pair in both sides of the final cuts as in \cite{NQS:2005,Nayak:2006fm}. If we neglect the term $\lim_{t\to \infty(1-i\epsilon)}
e^{-i(2E_{q}-M_{J/\psi}+2M_{c})t}$, which does not affect the cancellation of infrared divergences, then topologically unfactorized infrared divergent part of the summation of these diagrams reads(\cite{NQS:2005,Nayak:2006fm}):
\begin{equation}
\label{infd}
\epsilon^{(8\to 1)}(c\bar{c})=-\frac{N_{c}}{4}(N_{c}^{2}-1)
\frac{\alpha_{s}^{2}}{4\epsilon}[1-\frac{1}{f(|\vec{v}|)}\ln(\frac{1+f(|\vec{v}|)}{1-f(|\vec{v}|)})]+O(\alpha_{s}^{3})
\end{equation}
\begin{equation}
f(x)=\frac{2x}{1+x^{2}}
\end{equation}
, where $\vec{v}$ is the relative velocity of the heavy quark in the center- of-mass frame of the heavy quark pair.

\begin{figure*}
\begin{tabular}{c@{\hspace*{10mm}}c@{\hspace*{10mm}}c}
\includegraphics[scale=0.3]{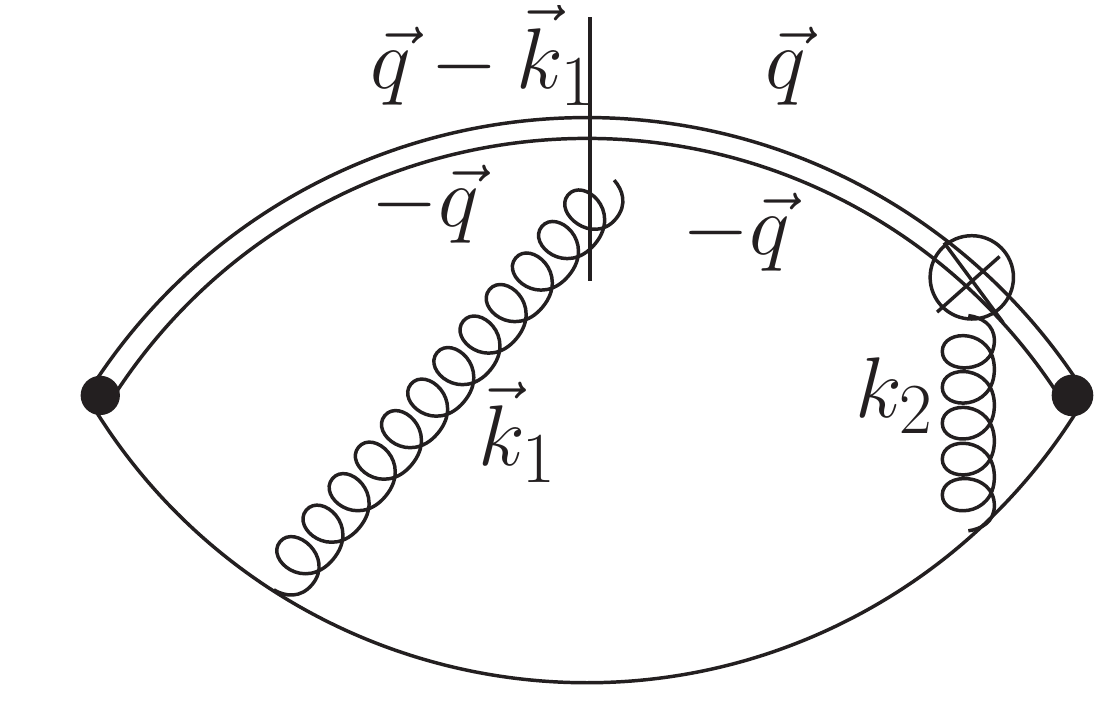}
&
\includegraphics[scale=0.3]{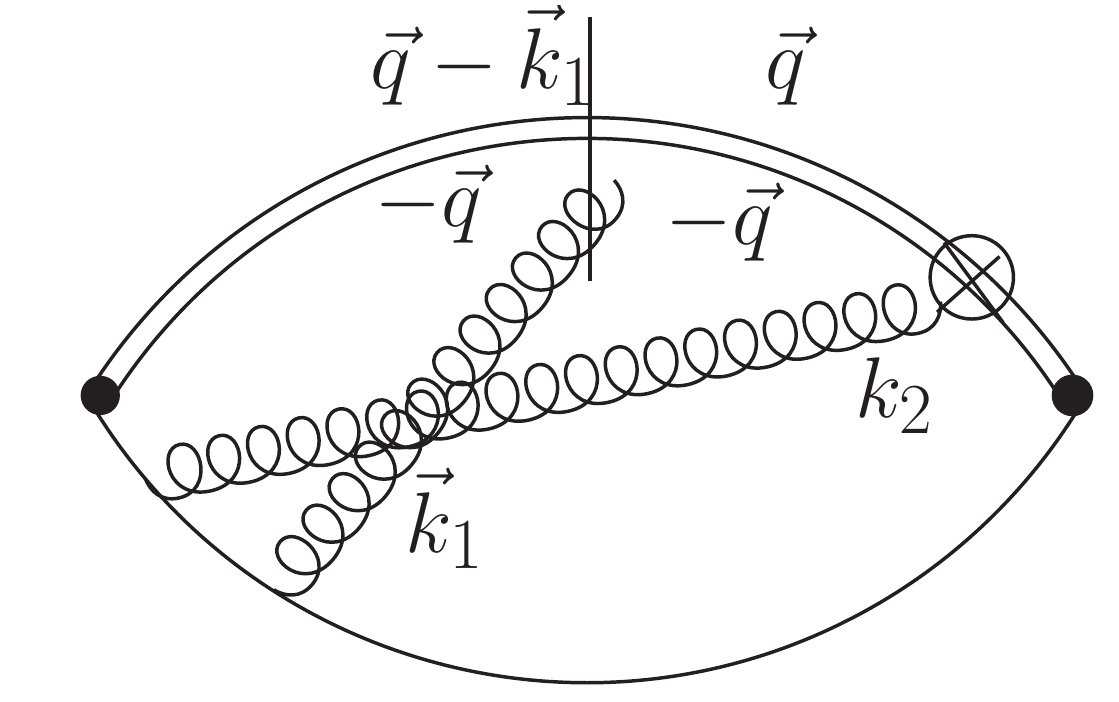}
&
\includegraphics[scale=0.3]{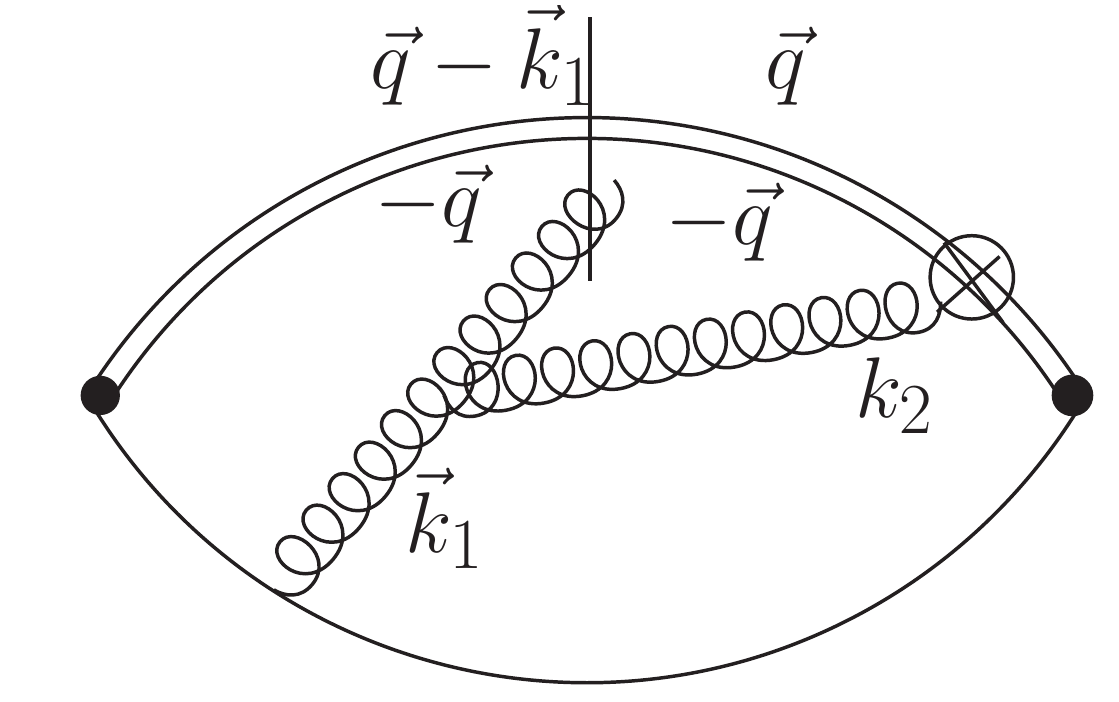}
\\
(a)&(b)&(c)
\\
\includegraphics[scale=0.3]{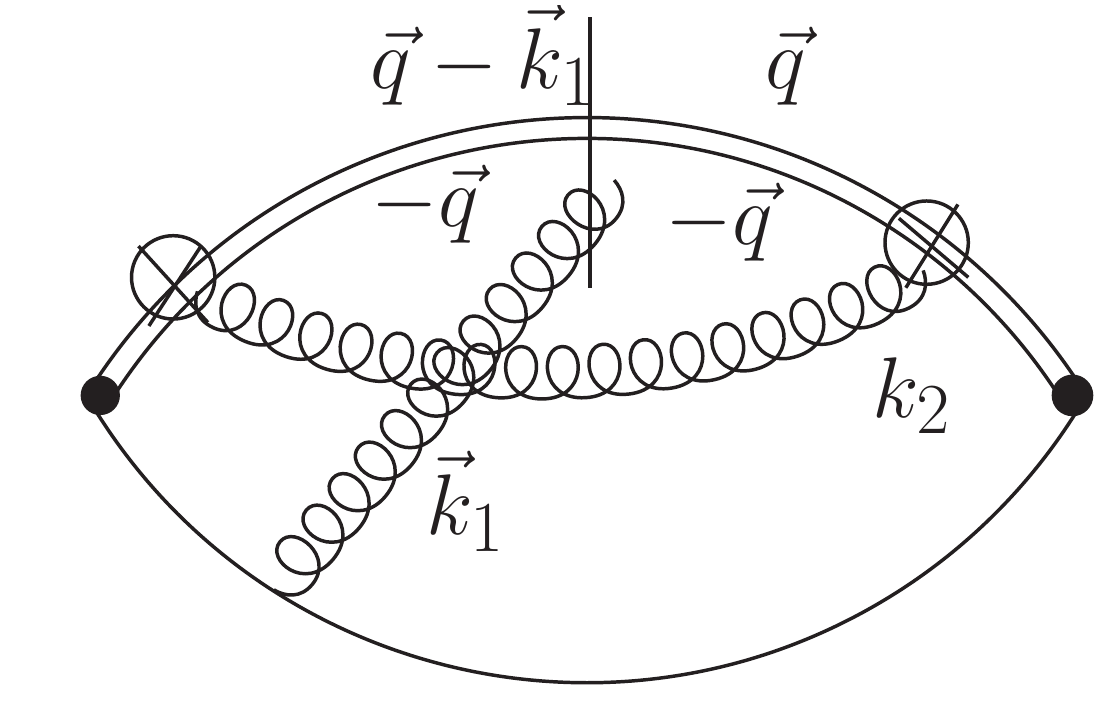}
&
\includegraphics[scale=0.3]{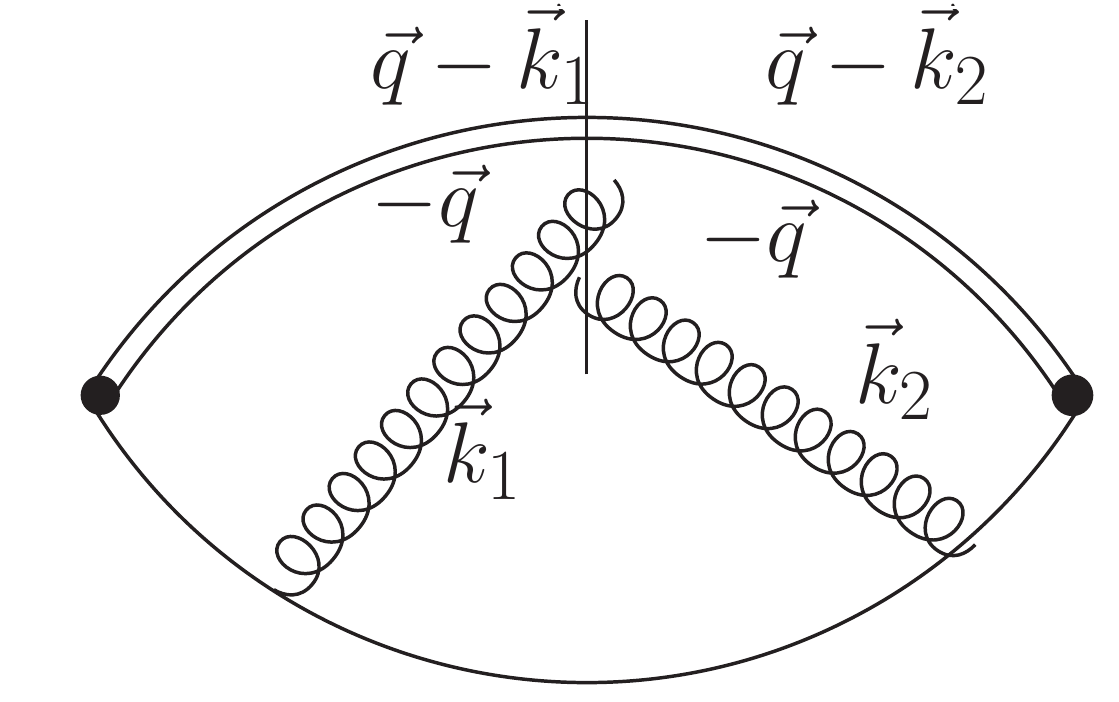}
\\
(d)&(e)
\end{tabular}
\caption{Diagrams with the Fock states $|c\bar{c}g\big>$ in the final states. }
\label{2fock}
\end{figure*}

\begin{figure*}
\begin{center}
\includegraphics[scale=0.5]{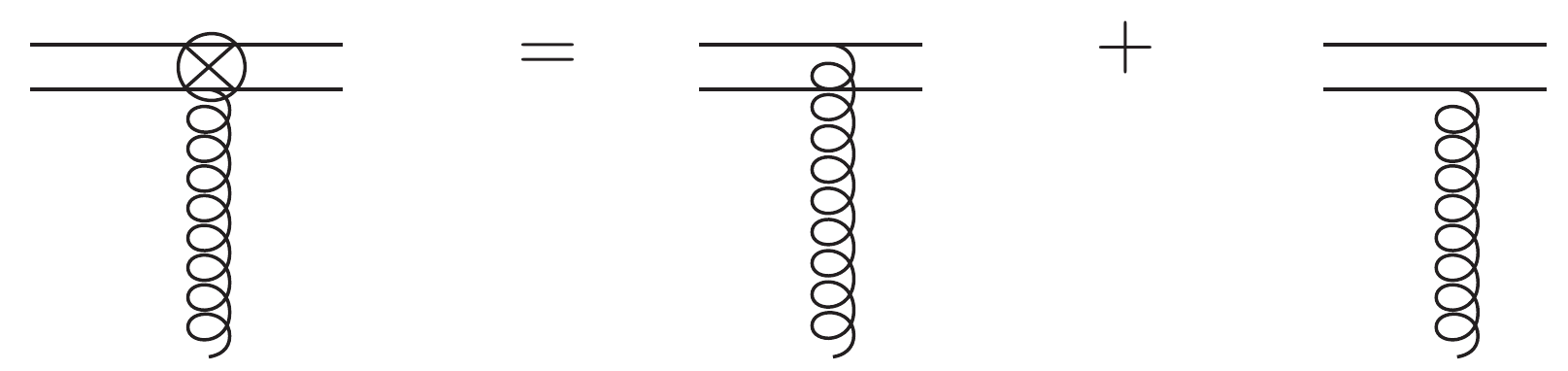}
\caption{Definition of the vertex $\bigotimes$ in Fig.\ref{2fock} }
\end{center}
\label{vertex}
\end{figure*}

Except for the lowest Fock state, the Fock state $|c\bar{c}g\big>$ in (\ref{jpsi-ccbar}) also contributes to the cross section up to NNLO in QCD interactions.  Diagrams with higher Fock states take the forms shown in Fig.\ref{2fock} or their conjugations, where the
effective vertex $\bigotimes$ is defined in Fig. \ref{vertex}. Color factors of the first and the last diagram in Fig.\ref{2fock} read:
\begin{equation}
C^{3a}\propto tr[t^{a}t^{b}]f^{abc}=0,\quad C^{3e}\propto tr[t^{a}t^{b}]tr[t^{d}t^{e}]f^{abc}f^{cde}=0
\end{equation}
.  We thus neglect these diagrams
and discuss remaining diagrams explicitly.
Without loss of generality, we take $l^{\mu}$ to be:
\begin{equation}
l^{\mu}=\frac{1}{\sqrt{2}}(1,0,0,1)
\end{equation}
in following calculations. Differences between these diagrams and those in Fig.\ref{1fock} are that real gluons in Fig.\ref{1fock}
belong to the undetected state $X$, while the real gluon $k_{1}$ in Fig.\ref{2fock} belongs to the constituents of the
detected $J/\psi$ particle.

\subsection{Diagrams that take the form shown in Fig.\ref{2fock}b}

\begin{figure*}
\begin{tabular}{c@{\hspace*{10mm}}c}
\includegraphics[scale=0.3]{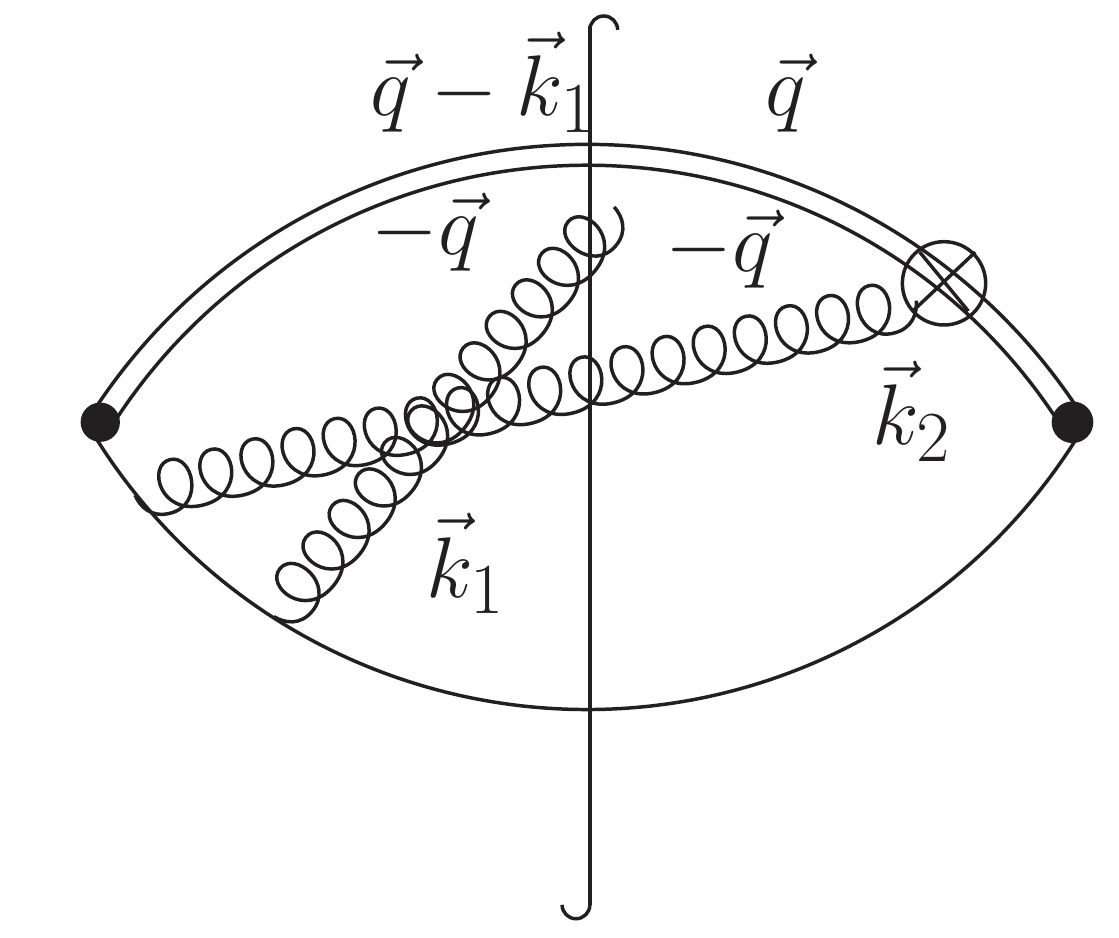}
&
\includegraphics[scale=0.3]{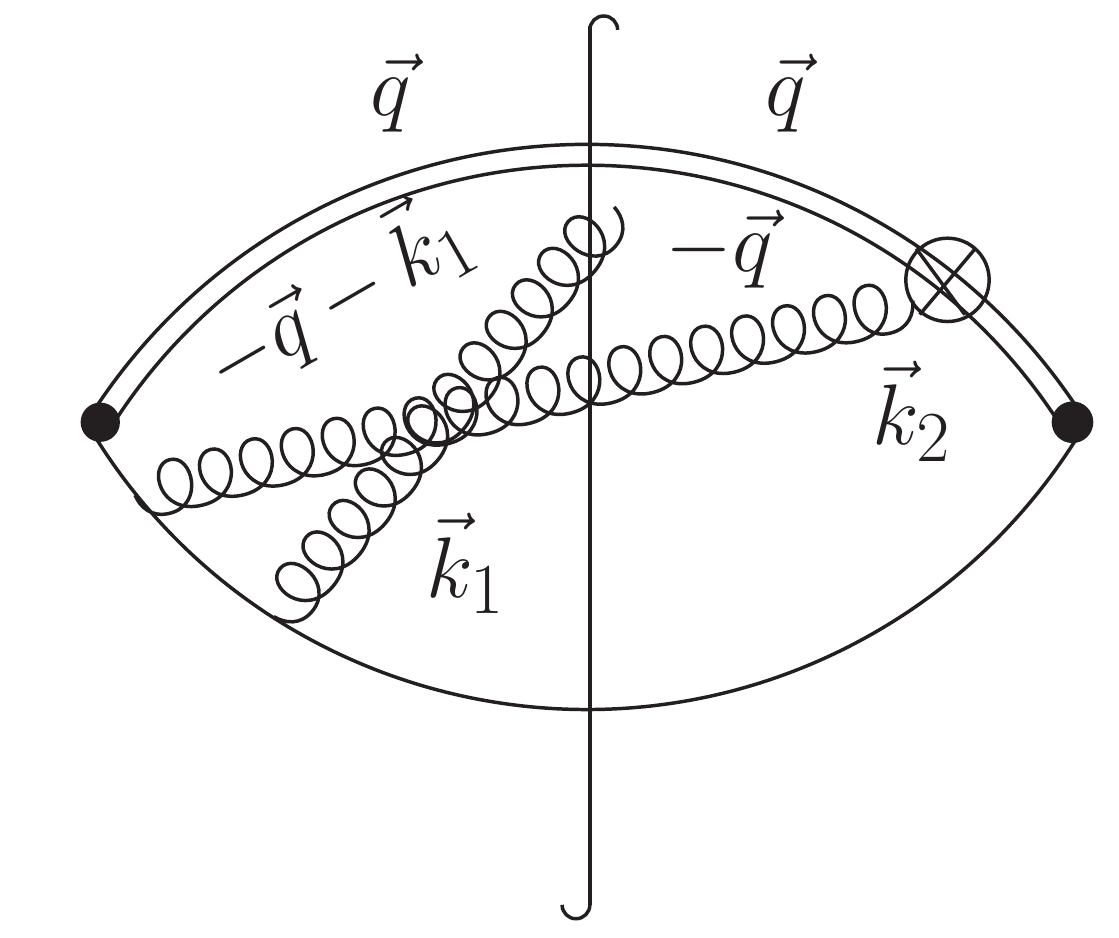}
\\
(a)&(b)
\end{tabular}
\caption{Diagrams that take the form shown  in Fig.\ref{2fock}b.}
\label{int1}
\end{figure*}

In this subsection, we consider the diagrams shown in Fig.\ref{int1}

For the first diagram in Fig.\ref{int1}, we have:
\begin{eqnarray}
\Sigma^{5a}(j)&=&\int\frac{\ud^{D-1} k_{1}}{(2\pi)^{D-1}2|\vec{k}_{1}|}\int\frac{\ud^{D-1}k_{2}}{(2\pi)^{D-1}2|\vec{k}_{2}|}
N^{5a}(j)\frac{1}{l\cdot k_{1}+i\epsilon}
\nonumber\\
&&
\frac{1}{l\cdot (k_{1}+k_{2})+i\epsilon}
\frac{1}{v_{1}\cdot k_{1}-i\epsilon}\frac{1}{v_{j}\cdot k_{2}-i\epsilon}
|_{\{k_{1}^{0}=|\vec{k}_{1}|,\quad k_{2}^{0}=|\vec{k}_{2}|\}}
\end{eqnarray}
, where $j=1,2$, the numerator term $N^{5a}(j)$ reads:
\begin{equation}
N^{5a}(j)=\left\{
   \begin{array}{ll}
    g^{4}\mu^{4\varepsilon}l\cdot v_{1}l\cdot v_{1}N_{c}(N_{c}^{2}-1)/4
    &\textrm{for $j=1$ }\\
    -g^{4}\mu^{4\varepsilon}l\cdot v_{1}l\cdot v_{2}N_{c}(N_{c}^{2}-1)/4
     &\textrm{for $j=2$ }
     \end{array}
     \right.
\end{equation}
. For the second diagram in Fig.\ref{int1}, we have:
\begin{eqnarray}
\Sigma^{5b}(j)&=&\int\frac{\ud^{D-1} k_{1}}{(2\pi)^{D-1}2|\vec{k}_{1}|}\int\frac{\ud^{D-1}k_{2}}{(2\pi)^{D-1}2|\vec{k}_{2}|}
N^{5b}(j)\frac{1}{l\cdot k_{1}+i\epsilon}
\nonumber\\
&&
\frac{1}{l\cdot (k_{1}+k_{2})+i\epsilon}
\frac{1}{v_{2}\cdot k_{1}-i\epsilon}\frac{1}{v_{j}\cdot k_{2}-i\epsilon}
|_{\{k_{1}^{0}=|\vec{k}_{1}|,\quad k_{2}^{0}=|\vec{k}_{2}|\}}
\end{eqnarray}
, where $j=1,2$, the numerator term $N^{5a}(j)$ reads:
\begin{equation}
N^{5b}(j)=\left\{
   \begin{array}{ll}
    -g^{4}\mu^{4\varepsilon}l\cdot v_{2}l\cdot v_{1}N_{c}(N_{c}^{2}-1)/4
    &\textrm{for $j=1$ }\\
    g^{4}\mu^{4\varepsilon}l\cdot v_{2}l\cdot v_{2}N_{c}(N_{c}^{2}-1)/4
     &\textrm{for $j=2$ }
     \end{array}
     \right.
\end{equation}
. The summation of these diagrams reads:
\begin{eqnarray}
&&\sum_{j=1}^{2}(\Sigma^{5a}(j)+\Sigma^{5b}(j))
\nonumber\\
&=&
g^{4}\mu^{4\varepsilon}\frac{N_{c}(N_{c}^{2}-1)}{4}
\int\frac{\ud^{D-1} k_{1}}{(2\pi)^{D-1}2|\vec{k}_{1}|}\int\frac{\ud^{D-1}k_{2}}{(2\pi)^{D-1}2|\vec{k}_{2}|}
\frac{1}{l\cdot k_{1}+i\epsilon}
\nonumber\\
&&
\frac{1}{l\cdot (k_{1}+k_{2})+i\epsilon}
(\frac{l\cdot v_{1}}{v_{1}\cdot k_{1}-i\epsilon}-\frac{l\cdot v_{2}}{v_{2}\cdot k_{1}-i\epsilon})
\nonumber\\
&&
(\frac{l\cdot v_{1}}{v_{1}\cdot k_{2}-i\epsilon}-\frac{l\cdot v_{2}}{v_{2}\cdot k_{2}-i\epsilon})
|_{\{k_{1}^{0}=|\vec{k}_{1}|,\quad k_{2}^{0}=|\vec{k}_{2}|\}}
\end{eqnarray}
If $k_{1}$ or $k_{2}$ is collinear to $l^{\mu}$, then we have:
\begin{equation}
v_{i}\cdot k_{j}=v_{i}^{-}k_{j}^{+}
\end{equation}
, where $i=1,2$, $j=1,2$.
One can verify that contributions of these regions cancel out between the diagrams shown in Fig.\ref{int1}.
We thus do not consider the collinear divergences of these diagrams here.

 We have:
 \begin{eqnarray}
&&\sum_{j=1}^{2}(\Sigma^{5a}(j)+\Sigma^{5b}(j))
\nonumber\\
&=&
g^{4}\mu^{4\varepsilon}\frac{N_{c}(N_{c}^{2}-1)}{4}
\int\frac{\ud^{D-2} k_{1\perp}}{(2\pi)^{D-2}}\int\frac{\ud^{D-2}k_{2\perp}}{(2\pi)^{D-2}}
\int_{0}^{\infty}\frac{\ud k_{1}^{-}}{2\pi}\int_{0}^{\infty}\frac{\ud k_{2}^{-}}{2\pi}
\nonumber\\
&&
\frac{1}{k_{1}^{-}+i\epsilon}
\frac{1}{k_{1}^{-}+k_{2}^{-}+i\epsilon}
\left(\frac{v_{1}^{-}}{2v_{1}^{+}(k_{1}^{-})^{2}+v_{1}^{-}k_{1\perp}^{2}-2k_{1}^{-}v_{1\perp}\cdot k_{1\perp}-i\epsilon}\right.
\nonumber\\
&&\left. -\frac{v_{2}^{-}}{2v_{2}^{+}(k_{1}^{-})^{2}+v_{2}^{-}k_{1\perp}^{2}-2k_{1}^{-}v_{2\perp}\cdot k_{1\perp}-i\epsilon}\right)
\nonumber\\
&&
\left(\frac{v_{1}^{-}}{2v_{1}^{+}(k_{2}^{-})^{2}+v_{1}^{-}k_{2\perp}^{2}-2k_{1}^{-}v_{1\perp}\cdot k_{2\perp}-i\epsilon}\right.
\nonumber\\
&&\left. -\frac{v_{2}^{-}}{2v_{2}^{+}(k_{2}^{-})^{2}+v_{2}^{-}k_{2\perp}^{2}-2k_{1}^{-}v_{2\perp}\cdot k_{2\perp}-i\epsilon}\right)
\nonumber\\
&=&\frac{\alpha_{s}^{\phantom{s}2}}{32\pi^{2}}
\frac{N_{c}(N_{c}^{2}-1)}{4}
(\frac{1}{\varepsilon})^{2}
[\ln^{2}\frac{(v_{1}\cdot l)^{2}}{(v_{2}\cdot l)^{2}}-2\varepsilon\ln\frac{\Lambda^{2}}{\mu^{2}}
\ln^{2}\frac{(v_{1}\cdot l)^{2}}{(v_{2}\cdot l)^{2}}
\nonumber\\
&&
+\varepsilon(\ln^{2}\frac{(v_{1}\cdot l)^{2}}{(v_{1})^{2}}
-\ln^{2}\frac{(v_{2}\cdot l)^{2}}{(v_{2})^{2}})\ln\frac{(v_{1}\cdot l)^{2}}{(v_{2}\cdot l)^{2}}]
\nonumber\\
&&
+(\text{infrared safe terms})
\end{eqnarray}
, where $\Lambda$ is the parameter chosen to regularize ultraviolet divergences.

For conjugations of diagrams in Fig.\ref{int1}, we have the same result. That is:
\begin{eqnarray}
\label{int1s}
&&(Fig.\ref{int1}+\text{conjugations})
\nonumber\\
&=&\frac{\alpha_{s}^{\phantom{s}2}}{16\pi^{2}}
\frac{N_{c}(N_{c}^{2}-1)}{4}
(\frac{1}{\varepsilon})^{2}
[\ln^{2}\frac{(v_{1}\cdot l)^{2}}{(v_{2}\cdot l)^{2}}-2\varepsilon\ln\frac{\Lambda^{2}}{\mu^{2}}
\ln^{2}\frac{(v_{1}\cdot l)^{2}}{(v_{2}\cdot l)^{2}}
\nonumber\\
&&
+\varepsilon(\ln^{2}\frac{(v_{1}\cdot l)^{2}}{(v_{1})^{2}}
-\ln^{2}\frac{(v_{2}\cdot l)^{2}}{(v_{2})^{2}})\ln\frac{(v_{1}\cdot l)^{2}}{(v_{2}\cdot l)^{2}}]
\nonumber\\
&&
+(\text{infrared safe terms})
\end{eqnarray}

\subsection{Diagrams that take the form shown in Fig.\ref{2fock}c}

\begin{figure*}
\begin{tabular}{c@{\hspace*{10mm}}c}
\includegraphics[scale=0.3]{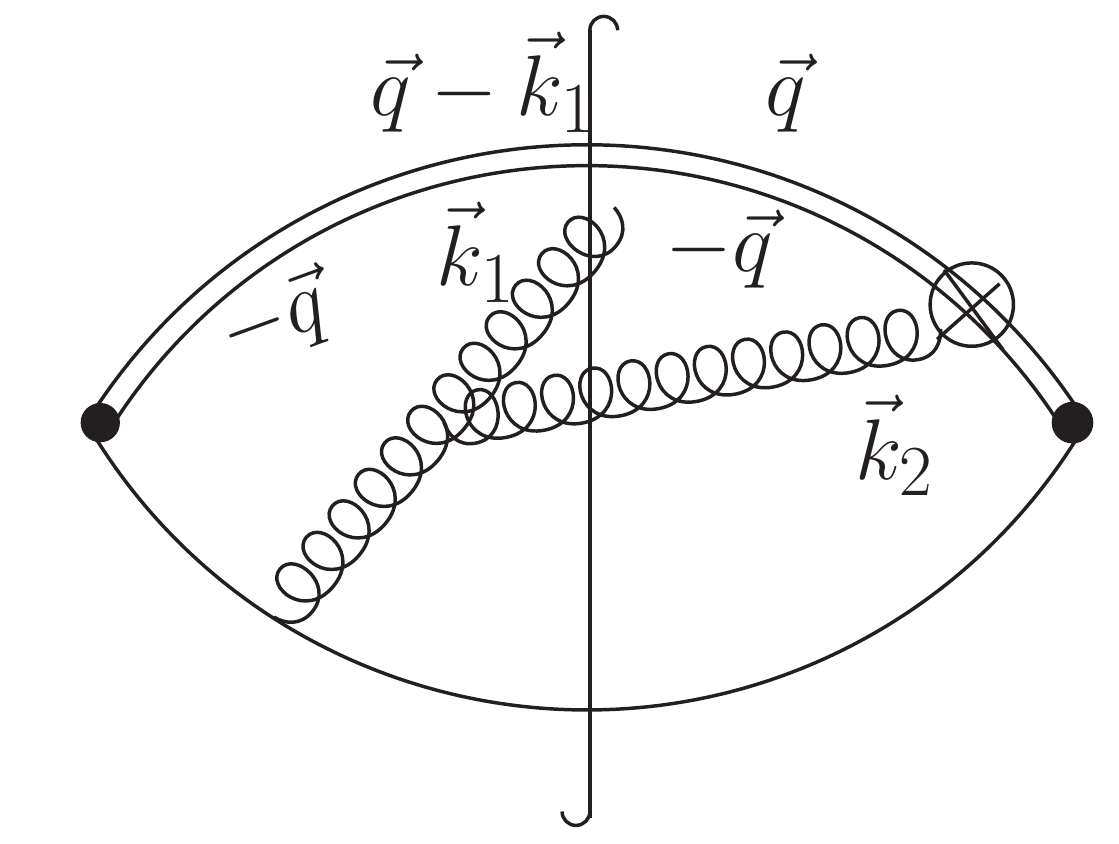}
&
\includegraphics[scale=0.3]{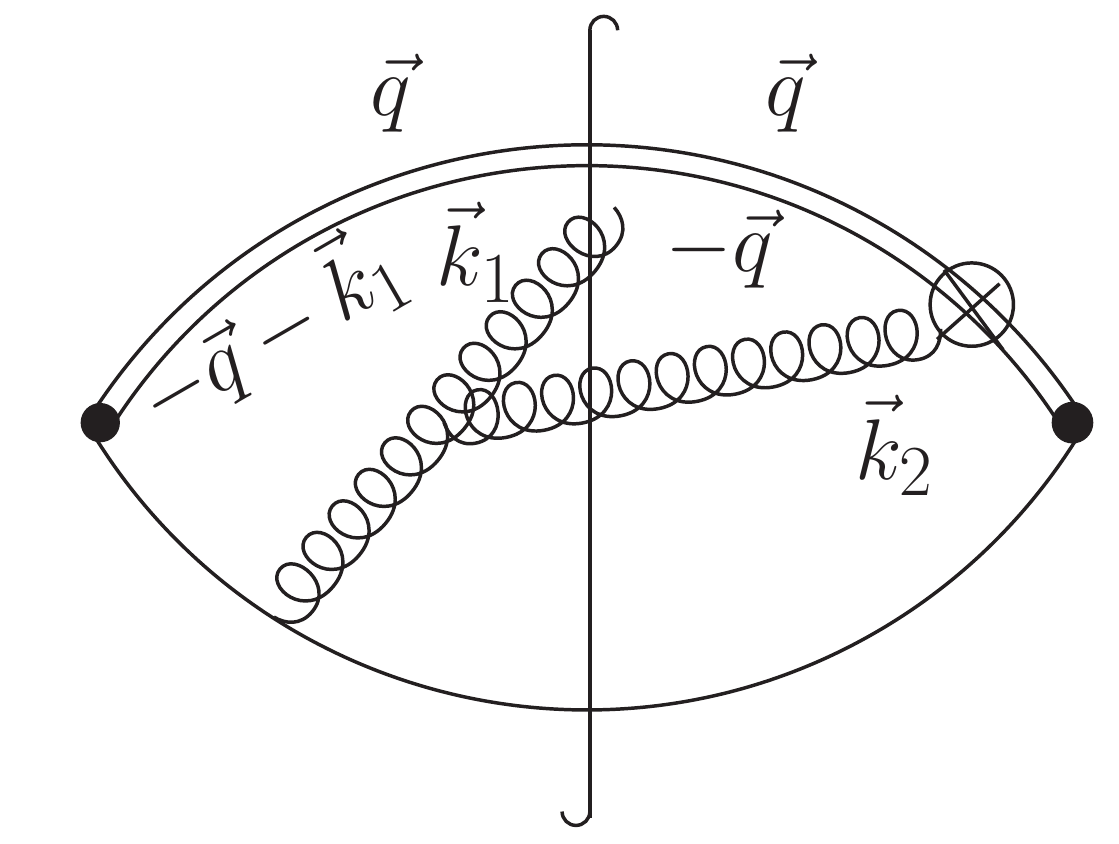}
\\
(a)&(b)
\end{tabular}
\caption{Diagrams that take the form shown  in Fig.\ref{2fock}c.}
\label{int2}
\end{figure*}

In this subsection, we consider diagrams shown in Fig.\ref{int2}.

For the first diagram in Fig.\ref{int2}, we have:
\begin{eqnarray}
\Sigma^{6a}(j)&=&\int\frac{\ud^{D-1} k_{1}}{(2\pi)^{D-1}2|\vec{k}_{1}|}\int\frac{\ud^{D-1}k_{2}}{(2\pi)^{D-1}2|\vec{k}_{2}|}
N^{6a}(k_{1},k_{2},j)\frac{1}{(k_{1}+k_{2})^{2}+i\epsilon}
\nonumber\\
&&
\frac{1}{l\cdot (k_{1}+k_{2})+i\epsilon}
\frac{1}{v_{1}\cdot k_{1}-i\epsilon}\frac{1}{v_{j}\cdot k_{2}-i\epsilon}
|_{\{k_{1}^{0}=|\vec{k}_{1}|,\quad k_{2}^{0}=|\vec{k}_{2}|\}}
\end{eqnarray}
, where $j=1,2$, the numerator term $N^{6a}(j)$ reads:
\begin{equation}
N^{6a}(k_{1},k_{2},j)=\left\{
   \begin{array}{ll}
    \frac{g^{4}\mu^{4\varepsilon}}{4}N_{c}(N_{c}^{2}-1)
    [v_{1}\cdot l v_{1}\cdot (-2k_{1}-k_{2})+v_{1}\cdot l v_{1}\cdot (k_{1}+2k_{2})+v_{1}\cdot v_{1} l\cdot (k_{1}-k_{2})]
    &\textrm{for $j=1$ }\\
    -\frac{g^{4}\mu^{4\varepsilon}}{4}N_{c}(N_{c}^{2}-1)
    [v_{1}\cdot l v_{2}\cdot (-2k_{1}-k_{2})+v_{2}\cdot l v_{1}\cdot (k_{1}+2k_{2})+v_{1}\cdot v_{2} l\cdot (k_{1}-k_{2})]
     &\textrm{for $j=2$ }
     \end{array}
     \right.
\end{equation}
.
For the second diagram in Fig.\ref{int2}, we have:
\begin{eqnarray}
\Sigma^{6b}(j)&=&\int\frac{\ud^{D-1} k_{1}}{(2\pi)^{D-1}2|\vec{k}_{1}|}\int\frac{\ud^{D-1}k_{2}}{(2\pi)^{D-1}2|\vec{k}_{2}|}
N^{6b}(k_{1},k_{2},j)\frac{1}{(k_{1}+k_{2})^{2}+i\epsilon}
\nonumber\\
&&
\frac{1}{l\cdot (k_{1}+k_{2})+i\epsilon}
\frac{1}{v_{2}\cdot k_{1}-i\epsilon}\frac{1}{v_{j}\cdot k_{2}-i\epsilon}
|_{\{k_{1}^{0}=|\vec{k}_{1}|,\quad k_{2}^{0}=|\vec{k}_{2}|\}}
\end{eqnarray}
, where $j=1,2$, the numerator term $N^{6b}(j)$ reads:
\begin{equation}
N^{6b}(k_{1},k_{2},j)=\left\{
   \begin{array}{ll}
    -\frac{g^{4}\mu^{4\varepsilon}}{4}N_{c}(N_{c}^{2}-1)
    [v_{2}\cdot l v_{1}\cdot (-2k_{1}-k_{2})+v_{1}\cdot l v_{2}\cdot (k_{1}+2k_{2})+v_{2}\cdot v_{1} l\cdot (k_{1}-k_{2})]
    &\textrm{for $j=1$ }\\
    \frac{g^{4}\mu^{4\varepsilon}}{4}N_{c}(N_{c}^{2}-1)
    [v_{2}\cdot l v_{2}\cdot (-2k_{1}-k_{2})+v_{2}\cdot l v_{2}\cdot (k_{1}+2k_{2})+v_{2}\cdot v_{2} l\cdot (k_{1}-k_{2})]
     &\textrm{for $j=2$ }
     \end{array}
     \right.
\end{equation}
.

We first consider contributions of collinear regions. Contributions of the region that $k_{1}$($k_{2}$) is collinear to $l^{\mu}$ and $l\cdot k_{2}$($l\cdot k_{1}$) is finite are power suppressed. In the region that $k_{1}$($k_{2}$) is collinear to $l^{\mu}$ and $k_{2}$($k_{1}$) is infrared, we have:
\begin{equation}
v_{1}\cdot k_{1}=v_{1}^{-}k_{1}^{+},\quad M^{6a}(k_{1},k_{2},j)=M^{6a}(\widetilde{k}_{1},0,j)
\end{equation}
or
\begin{equation}
v_{j}\cdot k_{2}=v_{j}^{-}k_{2}^{+},\quad M^{6a}(k_{1},k_{2},j)=M^{6a}(0,\widetilde{k}_{2},j)
\end{equation}
, where
\begin{equation}
(\widetilde{k}_{1}^{+},\widetilde{k}_{1}^{-},\vec{\widetilde{k}}_{1\perp})=(k_{1}^{+},0,\vec{0})
\end{equation}
\begin{equation}
(\widetilde{k}_{2}^{+},\widetilde{k}_{2}^{-},\vec{\widetilde{k}}_{2\perp})=(k_{2}^{+},0,\vec{0})
\end{equation}
. One can verbify that contributions of these region cancel out between diagrams shown in Fig.\ref{int2}. If both $k_{1}$ and $k_{2}$ are collinear to $l^{\mu}$, then we have:
\begin{equation}
v_{1}\cdot k_{1}=v_{1}^{-}k_{1}^{+},\quad
v_{j}\cdot k_{2}=v_{j}^{-}k_{2}^{+}
\end{equation}
. Contributions of such region cancel out between $\Sigma^{6a}(1)$ and $\Sigma^{6a}(2)$.
If $k_{1}$ is collinear to $k_{2}$ with $l\cdot k_{1}$ and $l\cdot k_{2}$ finite, then we have:
\begin{equation}
v_{1}\cdot k_{1}\propto v_{1}\cdot k_{2}, \quad v_{j}\cdot k_{1}\propto v_{2}\cdot k_{2}, \quad l\cdot k_{1} \propto l\cdot k_{2}
\end{equation}
Contributions of such region also cancel out between $\Sigma^{6a}(1)$ and $\Sigma^{6a}(2)$.
Thus collinear divergences do not disturb us.

The summation of $\Sigma^{6a}(j)$ and $\Sigma^{6b}(j)$ vanishes as integrands  are antisymmetric
under the the exchange $\vec{k_{1}} \leftrightarrow \vec{k_{2}}$. We see that infrared divergent terms in diagrams shown in Fig.\ref{int2} cancel out.
We have:
\begin{equation}
\label{int2s}
Fig.\ref{int2}=(Fig.\ref{int2})^{*}=0
\end{equation}

\subsection{Diagrams that take the form shown in Fig.\ref{2fock}d}

\begin{figure*}
\begin{tabular}{c@{\hspace*{10mm}}c}
\includegraphics[scale=0.3]{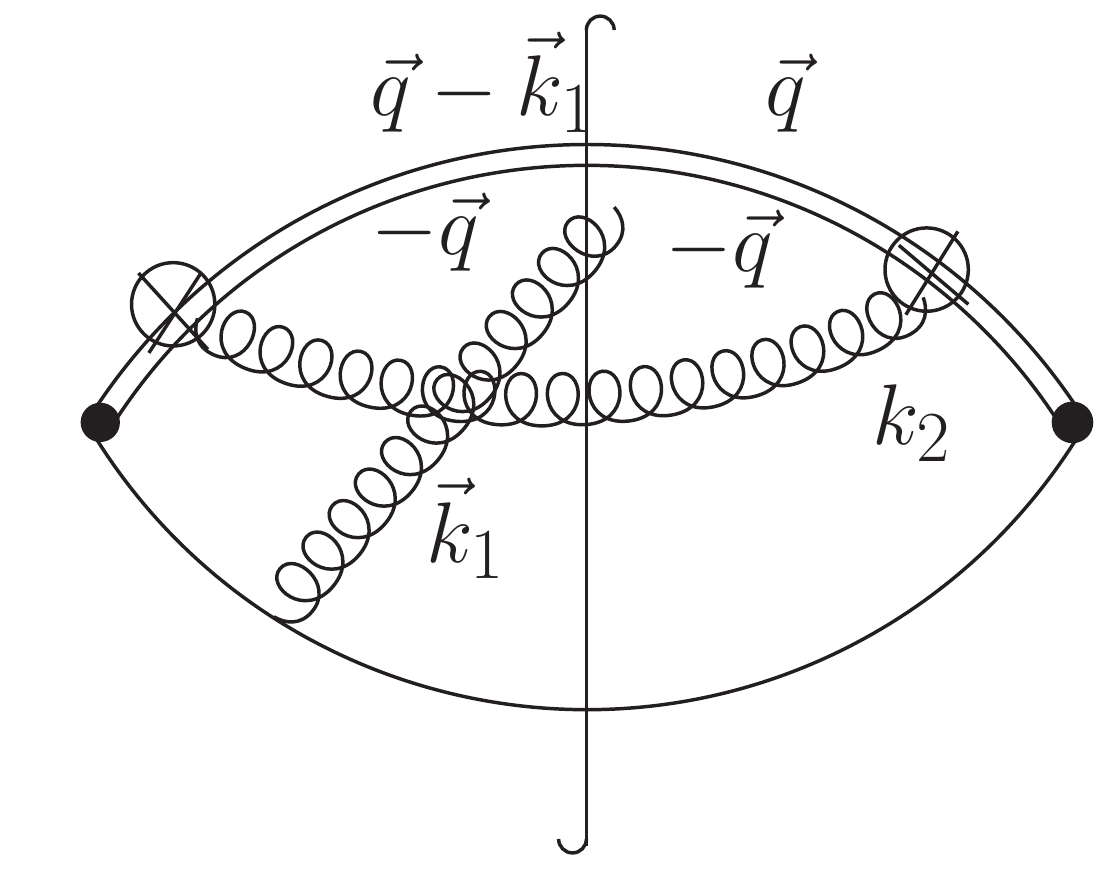}
&
\includegraphics[scale=0.3]{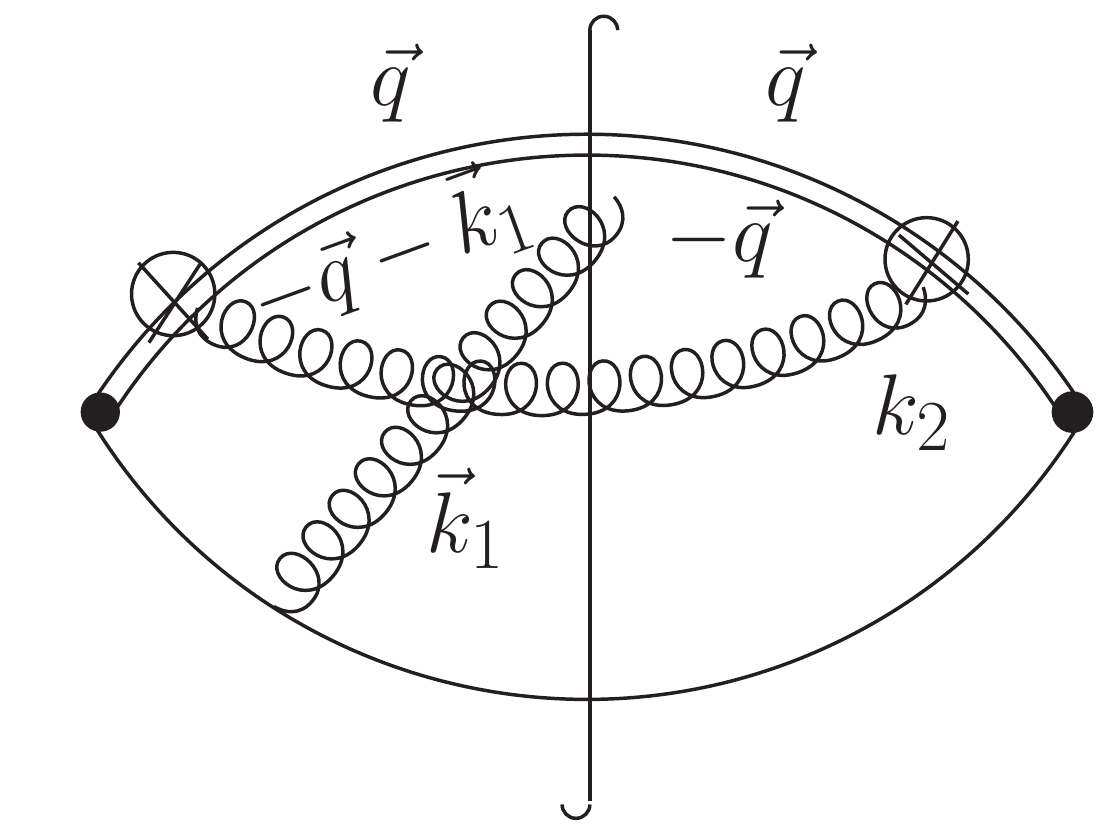}
\\
(a)&(b)
\end{tabular}
\caption{Diagrams that take the form shown  in Fig.\ref{2fock}d.}
\label{int3}
\end{figure*}

In this subsection, we consider the diagrams shown in Fig.\ref{int3}.
For the first diagram in Fig.\ref{int3}, we have:
\begin{eqnarray}
\Sigma^{7a}(j,j^{\prime})&=&
\int\frac{\ud^{D-1}k_{1}}{(2\pi)^{D-1}2|\vec{k}_{1}|}
\int\frac{\ud^{D-1}k_{2}}{(2\pi)^{D-1}2|\vec{k}_{2}|}
N^{7a}(j,j^{\prime})
\nonumber\\
&&
\frac{1}{l\cdot k_{1}+i\epsilon}
\frac{1}{v_{1}\cdot k_{1}-i\epsilon}\frac{1}{v_{j}\cdot k_{2}+i\epsilon}\frac{1}{v_{j^{\prime}}\cdot k_{2}-i\epsilon}
|_{\{k_{1}^{0}=|\vec{k}_{1}|,\quad k_{2}^{0}=|\vec{k}_{2}|\}}
\end{eqnarray}
, where $j=1,2$, $j^{\prime}=1,2$, the numerator term $N^{6a}(j,j^{\prime})$ is defined as:
\begin{equation}
N^{7a}(j,j^{\prime})=\left\{
   \begin{array}{ll}
    -\frac{g^{4}\mu^{4\varepsilon}}{8}N_{c}(N_{c}^{2}-1)v_{1}\cdot v_{1}l\cdot v_{1}
    &\textrm{for $j=j^{\prime}=1$}\\
     \frac{g^{4}\mu^{4\varepsilon}}{8}N_{c}(N_{c}^{2}-1)v_{1}\cdot v_{2}l\cdot v_{1}
     &\textrm{for $j=1$, $j^{\prime}=2$ }\\
     -\frac{g^{4}\mu^{4\varepsilon}}{8}N_{c}(N_{c}^{2}-1)v_{1}\cdot v_{2}l\cdot v_{1}
     &\textrm{for $j=2$, $j^{\prime}=1$ }\\
     \frac{g^{4}\mu^{4\varepsilon}}{8}N_{c}(N_{c}^{2}-1)v_{2}\cdot v_{2}l\cdot v_{1}
     &\textrm{for $j=j^{\prime}=2$ }
     \end{array}
     \right.
\end{equation}
For the second diagram in Fig.\ref{int3}, we have:
\begin{eqnarray}
\Sigma^{7b}(j,j^{\prime})&=&
\int\frac{\ud^{D-1}k_{1}}{(2\pi)^{D-1}2|\vec{k}_{1}|}
\int\frac{\ud^{D-1}k_{2}}{(2\pi)^{D-1}2|\vec{k}_{2}|}
N^{7b}(j,j^{\prime})
\nonumber\\
&&
\frac{1}{l\cdot k_{1}+i\epsilon}
\frac{1}{v_{2}\cdot k_{1}-i\epsilon}\frac{1}{v_{j}\cdot k_{2}+i\epsilon}\frac{1}{v_{j^{\prime}}\cdot k_{2}-i\epsilon}
|_{\{k_{1}^{0}=|\vec{k}_{1}|,\quad k_{2}^{0}=|\vec{k}_{2}|\}}
\end{eqnarray}
, where $j=1,2$, $j^{\prime}=1,2$, the numerator term $N^{6a}(j,j^{\prime})$ is defined as:
\begin{equation}
N^{7b}(j,j^{\prime})=\left\{
   \begin{array}{ll}
    \frac{g^{4}\mu^{4\varepsilon}}{8}N_{c}(N_{c}^{2}-1)v_{1}\cdot v_{1}l\cdot v_{2}
    &\textrm{for $j=j^{\prime}=1$}\\
    -\frac{g^{4}\mu^{4\varepsilon}}{8}N_{c}(N_{c}^{2}-1)v_{1}\cdot v_{2}l\cdot v_{2}
     &\textrm{for $j=1$, $j^{\prime}=2$ }\\
     \frac{g^{4}\mu^{4\varepsilon}}{8}N_{c}(N_{c}^{2}-1)v_{1}\cdot v_{2}l\cdot v_{2}
     &\textrm{for $j=2$, $j^{\prime}=1$ }\\
     -\frac{g^{4}\mu^{4\varepsilon}}{8}N_{c}(N_{c}^{2}-1)v_{2}\cdot v_{2}l\cdot v_{2}
     &\textrm{for $j=j^{\prime}=2$ }
     \end{array}
     \right.
\end{equation}
 If $k_{1}$ is collinear to $l^{\mu}$, then we have:
 \begin{equation}
 v_{1}\cdot k_{1}\simeq v_{1}^{-}l_{1}^{+},\quad v_{2}\cdot k_{1}\simeq v_{2}^{-}k_{2}^{+}
 \end{equation}
. Contributions of such region cancel out between $\Sigma^{7a}(j,j^{\prime})$ and $\Sigma^{7b}(j,j^{\prime})$. We thus do not consider effects of collinear divergences.

We see that:
\begin{equation}
\sum_{j,j^{\prime}}(\Sigma^{7a}(j,j^{\prime})+\Sigma^{7b}(j,j^{\prime}))=0
\end{equation}
as integrands are antisymmetric under the transformation $\vec{k_{2}} \to -\vec{k_{2}}$. Thus infrared divergent terms in diagrams shown in Fig.\ref{int3} cancel out. We have:
\begin{equation}
\label{int3s}
Fig.\ref{int3}=(Fig.\ref{int3})^{*}=0
\end{equation}
.

According to (\ref{int1s}), (\ref{int2s}) and (\ref{int3s}), we have:
\begin{eqnarray}
\label{infdhf}
&&\epsilon^{(8\to 1)}(\text{high Fock states})
\nonumber\\
&=&\frac{\alpha_{s}^{\phantom{s}2}}{16\pi^{2}}
\frac{N_{c}(N_{c}^{2}-1)}{4}
(\frac{1}{\varepsilon})^{2}
[\ln^{2}\frac{(v_{1}\cdot l)^{2}}{(v_{2}\cdot l)^{2}}-2\varepsilon\ln\frac{\Lambda^{2}}{\mu^{2}}
\ln^{2}\frac{(v_{1}\cdot l)^{2}}{(v_{2}\cdot l)^{2}}
\nonumber\\
&&
+\varepsilon(\ln^{2}\frac{(v_{1}\cdot l)^{2}}{(v_{1})^{2}}
-\ln^{2}\frac{(v_{2}\cdot l)^{2}}{(v_{2})^{2}})\ln\frac{(v_{1}\cdot l)^{2}}{(v_{2}\cdot l)^{2}}]
\nonumber\\
&&
+(\text{infrared safe terms})
\nonumber\\
&&+O(\alpha_{s}^{\phantom{s}3})
\end{eqnarray}
We see that the summation of (\ref{infd}) and(\ref{infdhf}) is not infrared safe. Especially, infrared divergent terms in (\ref{infd}) are of order $\frac{1}{\varepsilon}$, while those in (\ref{infdhf} are of order $\frac{1}{\varepsilon^{2}}$ or $\frac{1}{\varepsilon}$. We conclude that fragmentation functions $D_{i\to J/\psi}$ and $D_{c\bar{c}(\kappa)\to J/\psi}$ suffer from topologically unfactorized infrared divergences shown in (\ref{infd}) even contributions of higher Fock states are taken into account.

\section{Inclusive Production of Stable Particles Near the Threshold of Heavy Quarkonia}
\label{incpro}

In practical process, detected heavy quarkonia  are reconstructed from their decay products. Without loss of generality, we consider the process:
\begin{equation}
A+B\to \mu^{+}\mu^{-}(n,p_{H})+X
\end{equation}
in this section, where $A$ and $B$ represent initial particles, $X$ represents undetected final particles, $n$ represents the quantum number of the heavy quarkonium one concerned, the momentum $p_{H}$ of the detected  $\mu^{+}\mu^{-}$ pair fulfill the condition:
\begin{equation}
p_{H}^{2}\simeq M^{2}
\end{equation}
with $M$ the mass of the heavy quarkonium one concerned.  According to the collinear factorization theorem presented in \cite{Kang:2011mg,Kang:2011zza,Kang:2014tta,Fleming:2012wy,Fleming:2013qu}, we have:
\begin{eqnarray}
\label{cfmu}
&&\sum_{X}\ud \sigma_{A+B\to \mu^{+}\mu^{-}(n,p_{H})+X}
\nonumber\\
&=&\sum_{i,X}\ud \sigma_{A+B\to i+X}\otimes D_{i\to \mu^{+}\mu^{-}(n,p_{H})}
\nonumber\\
&&
+
\sum_{\kappa,X} \ud \sigma_{ A+B\to Q\bar{Q}(\kappa)+X}\otimes D_{Q\bar{Q}(\kappa)\to \mu^{+}\mu^{-}(n,p_{H})}
\nonumber\\
&&+\mathcal{O}(M^{4}/p_{T}^{4})
\nonumber\\
&&+\ldots
\end{eqnarray}
, where $D_{i\to \mu^{+}\mu^{-}(n,p_{H})}$ and $D_{Q\bar{Q}(\kappa)\to \mu^{+}\mu^{-}(n,p_{H})}$ represent fragmentation functions for $i$ and $Q\bar{Q}(\kappa)$ to the state $\mu^{+}\mu^{-}(n,p_{H})$ under the evolution of QCD and QED interactions, the ellipsis represents contributions of process with the $\mu^{+}\mu^{-}$ pair produced in the short distance(order $1/p_{T}$) subprocess.

\begin{figure*}
\begin{tabular}{c@{\hspace*{10mm}}c@{\hspace*{10mm}}c}
\includegraphics[scale=0.3]{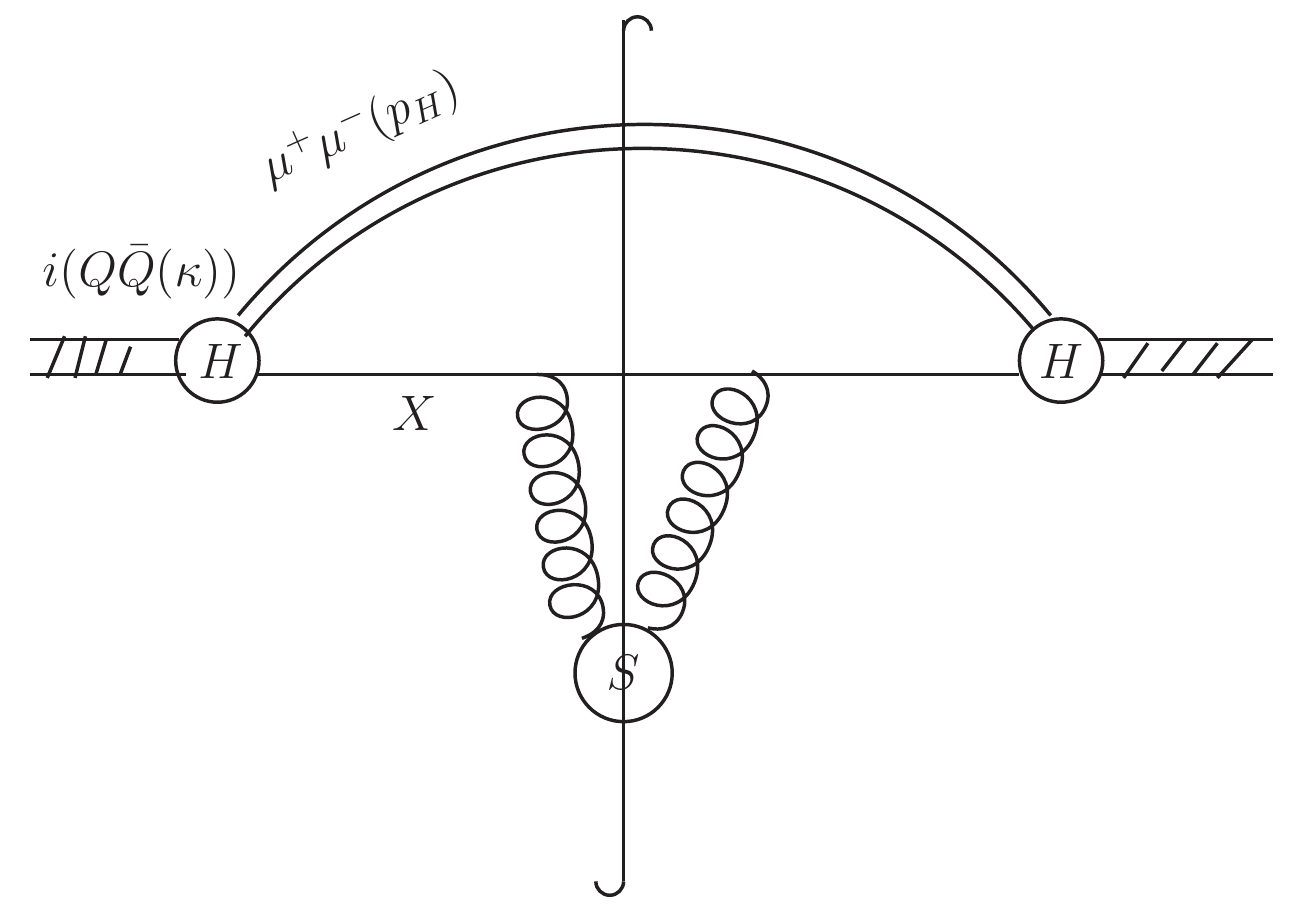}
&
\includegraphics[scale=0.3]{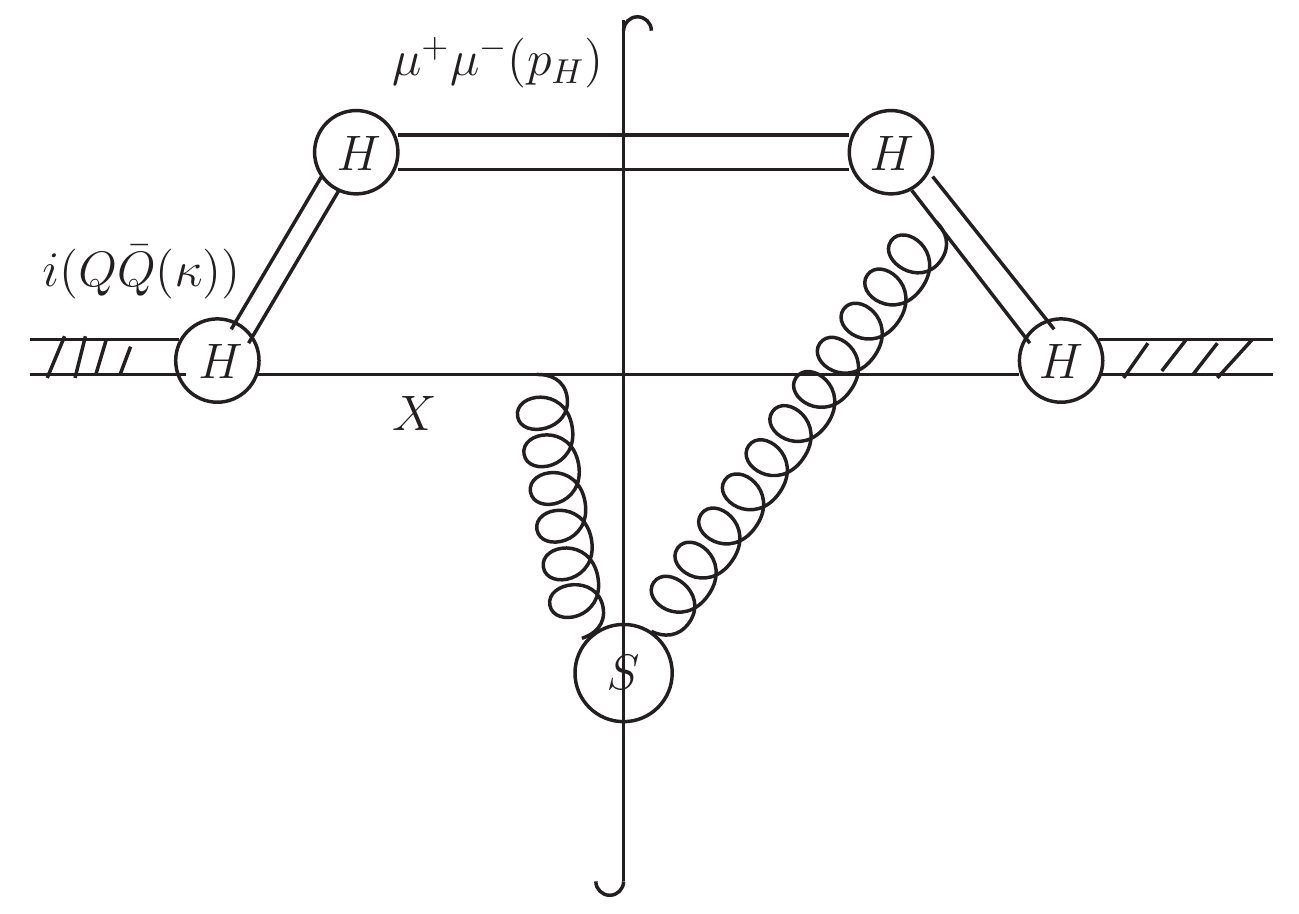}
&
\includegraphics[scale=0.3]{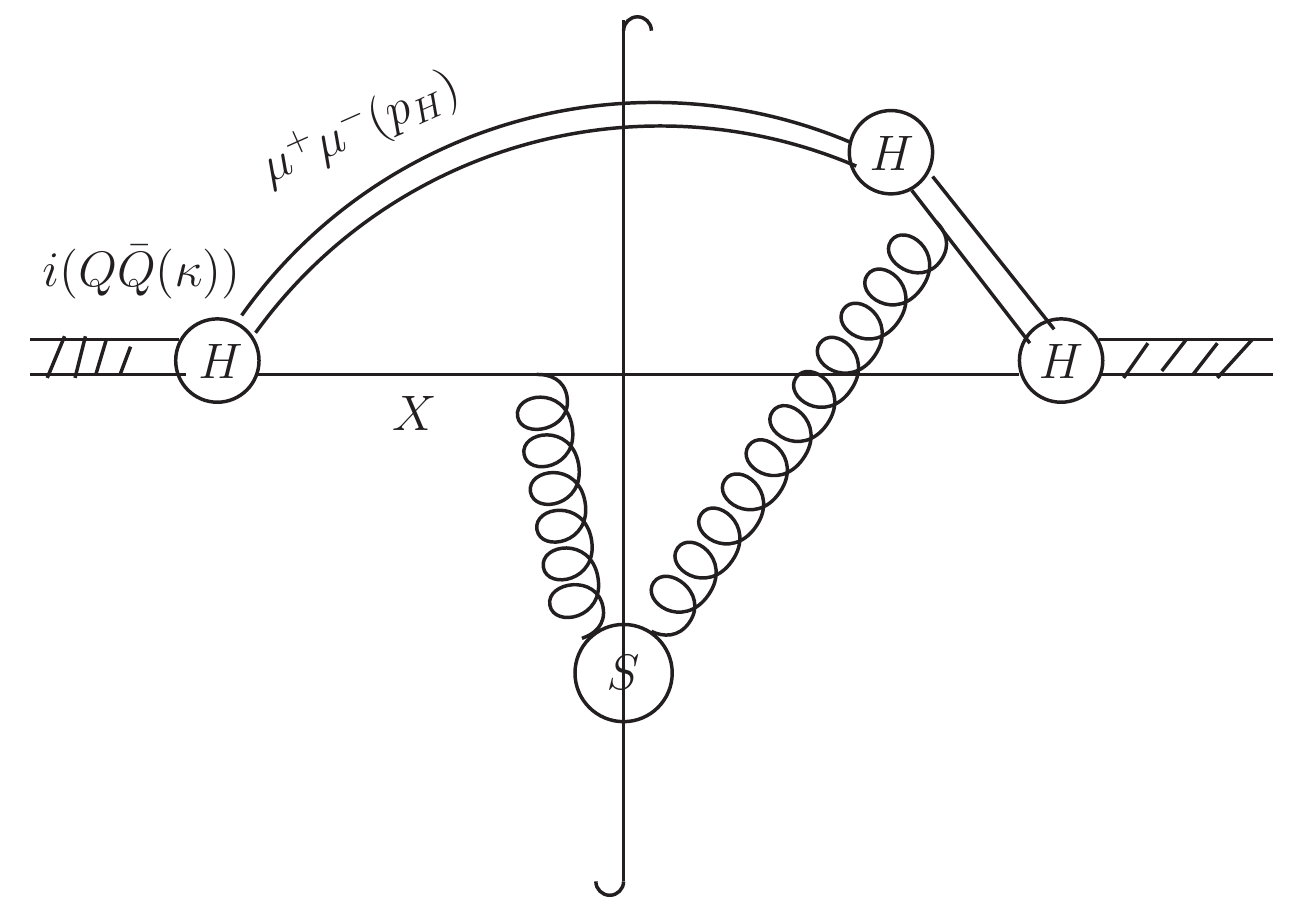}
\\
(a)&(b)&(c)
\end{tabular}
\caption{Figure \ref{incmu}:(a) Example of direct production of $\mu^{+}\mu^{-}$ pair in fragmentation functions $D_{i\to \mu^{+}\mu^{-}(n,p_{H})}$ and $D_{Q\bar{Q}(\kappa)\to \mu^{+}\mu^{-}(n,p_{H})}$; (b)Example of indirect production of $\mu^{+}\mu^{-}$ pair in fragmentation functions $D_{i\to \mu^{+}\mu^{-}(n,p_{H})}$ and $D_{Q\bar{Q}(\kappa)\to \mu^{+}\mu^{-}(n,p_{H})}$; (c) Example of interference terms between the direct and indirect production  of $\mu^{+}\mu^{-}$ pair.}
\label{incmu}
\end{figure*}

It is convenient to consider fragmentation functions $D_{i\to \mu^{+}\mu^{-}(n,p_{H})}$ and $D_{Q\bar{Q}(\kappa)\to \mu^{+}\mu^{-}(n,p_{H})}$ in the rest frame of the $\mu^{+}\mu^{-}$ pair. The $\mu^{+}\mu^{-}$ pair can be produced in a  direct short distance process(order $1/M$) involving the states $i$ or $Q\bar{Q}(\kappa)$. They can also be produced in a indirect short distance process(order $1/M$) involving intermediate states with momenta square of order $M^{2}$. We show examples of these two cases in Fig.\ref{incmu}.

Infrared divergences in fragmentation functions $D_{i\to \mu^{+}\mu^{-}(n,p_{H})}$ and $D_{Q\bar{Q}(\kappa)\to \mu^{+}\mu^{-}(n,p_{H})}$ are caused by: (1)infrared gluons exchanged between final undetected particles; (2)infrared gluons exchanged between final undetected particles and intermediate states that evolve to the $\mu^{+}\mu^{-}$ pair; (3)infrared gluons exchanged between intermediate states that evolve to the $\mu^{+}\mu^{-}$ pair;(4)infrared QED interactions and possible interference terms.

\begin{figure*}
\begin{tabular}{c@{\hspace*{10mm}}c}
\includegraphics[scale=0.5]{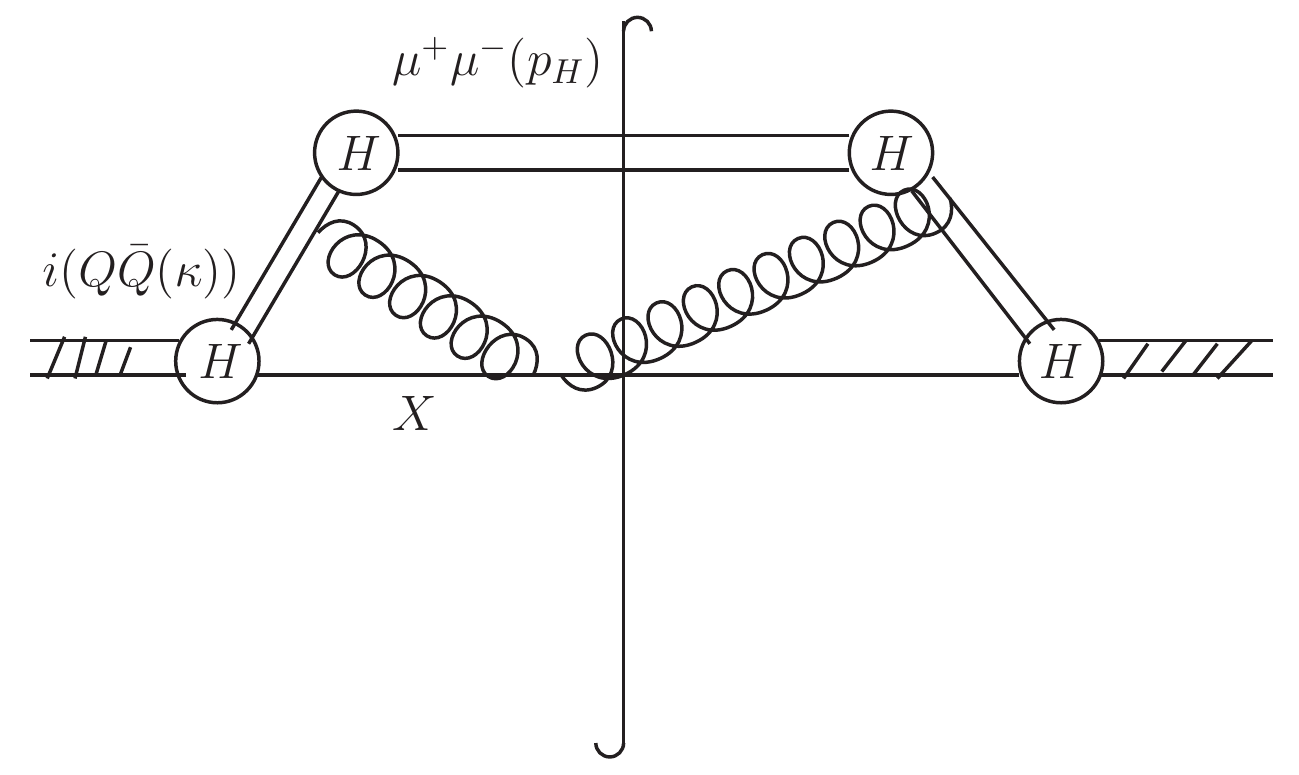}
&
\includegraphics[scale=0.5]{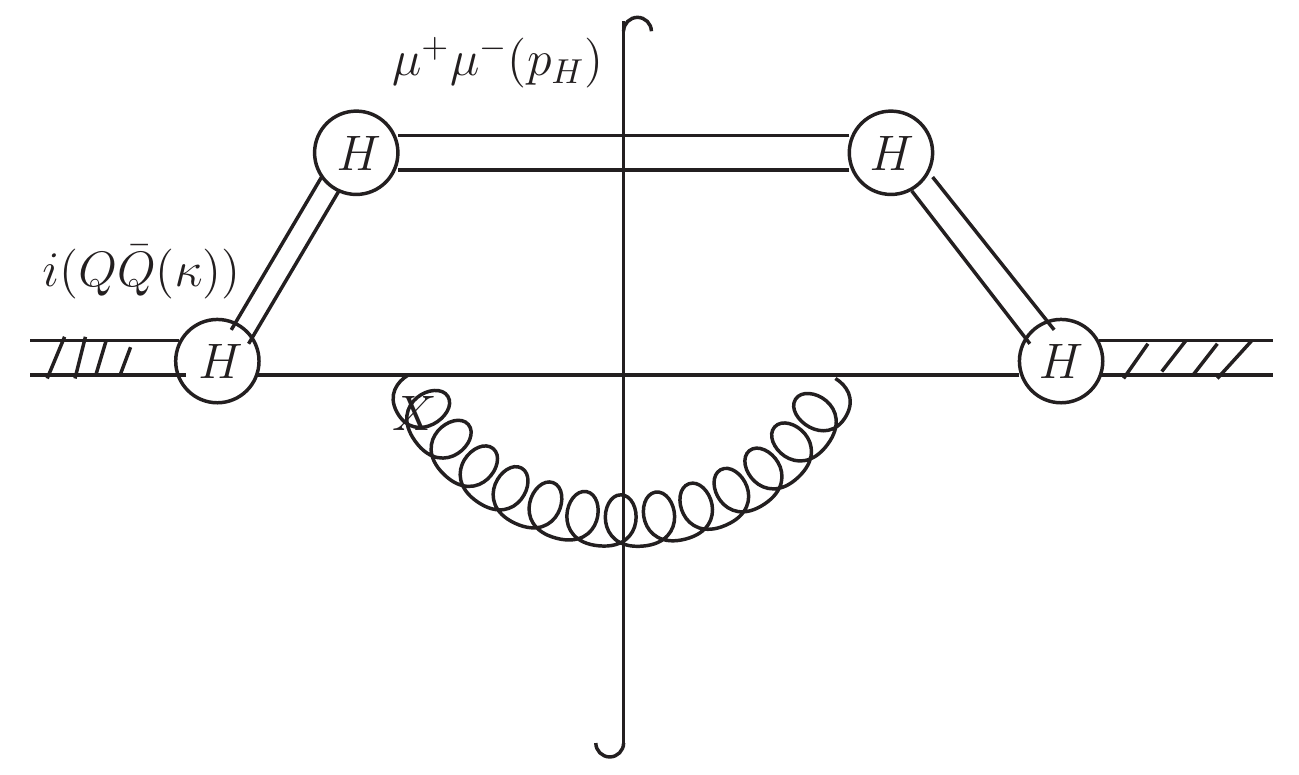}
\\
(a)&(b)
\end{tabular}
\caption{Figure \ref{infdiv}:(a) Example of topologically unfactorized QCD interactions;  (b)Example of topologically factorized QCD interactions.}
\label{infdiv}
\end{figure*}

We do not consider infrared QED interactions here, as such interactions are suppressed by the QED coupling constant. As in \cite{NQS:2005,Nayak:2006fm}, we classify the QCD interactions into two types: the topologically factorized and topologically unfactorized interactions. Examples of these two types are shown in Fig.\ref{infdiv}.

\begin{figure*}
\begin{center}
\includegraphics[scale=0.3]{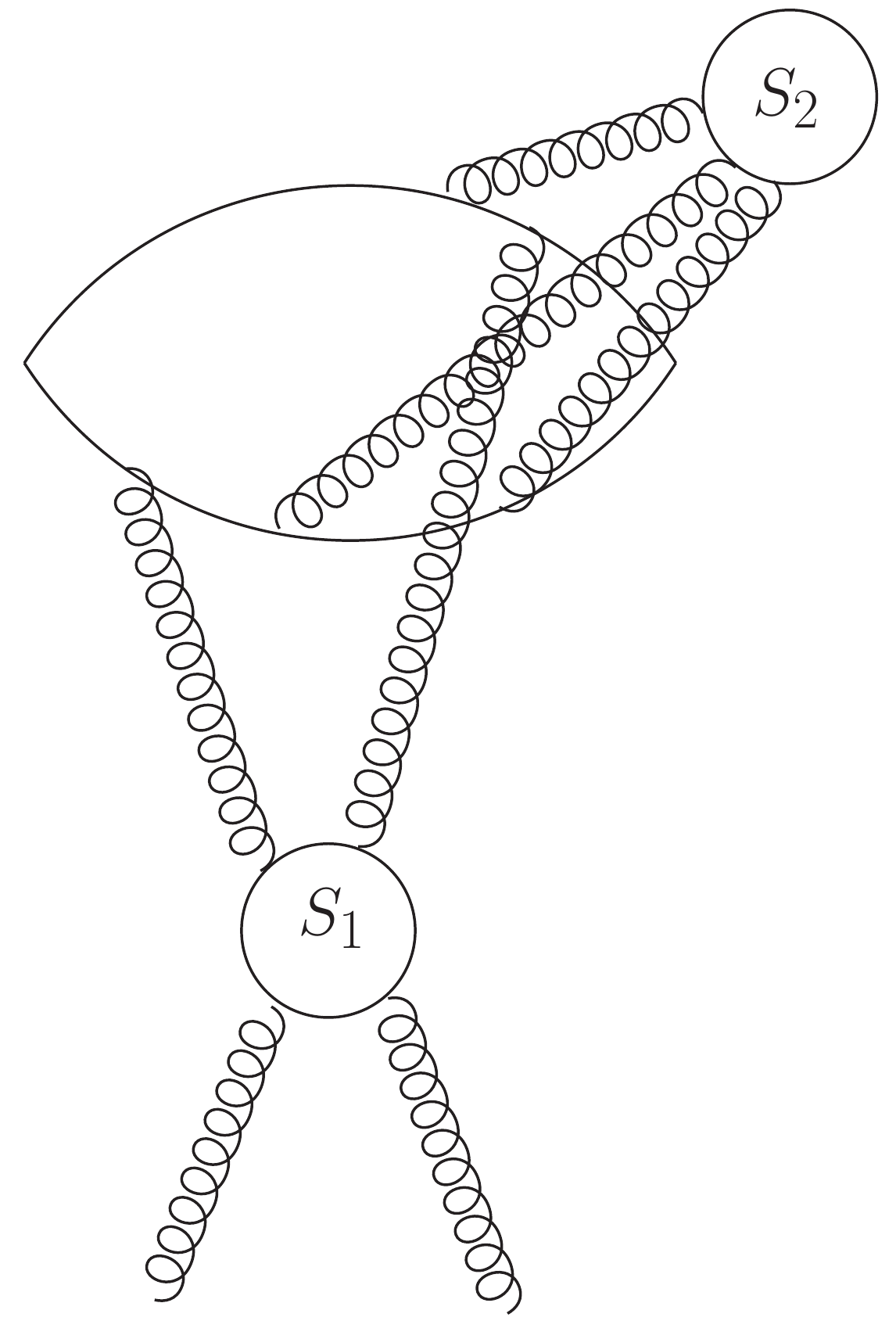}
\caption{Typical diagram that contributes to the infrared divergences of topologically unfactorized  diagrams}
\label{selfenergy}
\end{center}
\end{figure*}

For the first diagram in Fig.\ref{infdiv}, the infrared gluons do not produce pinch singular points unless the two intermediate quarks that annihilate to the $\mu^{+}\mu^{-}$ pair are at rest(\cite{CN:1965,S:1993}). In this case, effects of couplings between infrared gluons and   intermediate quarks can be absorbed into two Wilson lines that along the same (time-like)direction. We notice that
\begin{equation}
Y^{\dag}_{ij}(x)Y_{jk}(x)=\delta_{ik},\quad (Y)^{\dag}_{ij}(x)Y_{ki}(x)=\delta_{jk}
\end{equation}
for classical fields $A^{\mu}(x)$,  where $Y_{n}(x)$ is the  time-like Wilson line:
\begin{equation}
\label{unit}
Y(x)=\mathcal{P}\exp(-ig \int_{0}^{\infty}\ud s  A^{0}(x^{0}+s,\vec{x}))
\end{equation}
. To clarify the effects of self-energy graphs of Wilson lines, we consider the
 diagram shown in Fig.\ref{selfenergy}.
It can then be written as:
\begin{eqnarray}
\Sigma^{\mu\nu}_{10}(x,y,\mathcal{O}(A))&\equiv&
\big<T\{\bar{\psi}Y^{\dag}\gamma^{\mu}\mathcal{K}Y\psi(x)\mathcal{O}(A)\bar{\psi}Y^{\dag}\gamma^{\nu}Y\psi(y)\}\big>
\nonumber\\
&\propto&
\int \mathcal{D}\psi(z)\mathcal{D}\bar{\psi}(z)\mathcal{D} A^{\mu}(z)\bar{\psi}Y^{\dag}\gamma^{\mu}\mathcal{K}Y\psi(x)
\bar{\psi}Y^{\dag}\gamma^{\nu}Y\psi
\mathcal{O}(A)e^{i\int\ud^{4}z\mathcal{L}(\psi,A)(z)}
\end{eqnarray}
, where $\mathcal{K}=1$ for quarks produced as color singlet and $\mathcal{K}=t^{a}$ for that of color octet, $\bar{\psi}\gamma^{\nu}\psi$ is the electromagnetic current, $\mathcal{O}(A)$ represents operators at the other end of soft gluons of the type $S_{1}$,  the Lagrangian density $\mathcal{L}(\psi,A)(z)$ reads:
\begin{eqnarray}
\mathcal{L}(\psi,A)(z)&=&
\bar{\psi}(i\not\!\partial-m)\psi(z)
+\frac{1}{2g^{2}} tr ([\partial^{\mu}-igA^{\mu},\partial^{\nu}-igA^{\nu}])^{2}(z)
\end{eqnarray}
. We do not consider effects of gauge fixing term and ghost fields for simplicity. We have:
\begin{eqnarray}
\Sigma^{\mu\nu}_{10}(x,y,\mathcal{O}(A))&=&
\frac{\big<T\{\bar{\psi}Y^{\dag}\gamma^{\mu}Y\psi(x)\bar{\psi}Y^{\dag}\gamma^{\nu}Y\psi(y)\}\big>}
{\int [\mathcal{D}\psi(z)][\mathcal{D}\bar{\psi}(z)][\mathcal{D} A^{\mu}(z)]\bar{\psi}\gamma^{\mu}\psi(x)\bar{\psi}\gamma^{\nu}\psi
e^{i\int\ud^{4}z\mathcal{L}(\psi,A)(z)}}
\nonumber\\
&&
\int [\mathcal{D}\psi(z)][\mathcal{D}\bar{\psi}(z)][\mathcal{D} A^{\mu}(z)]\bar{\psi}\gamma^{\mu}\mathcal{K}\psi(x)\bar{\psi}\gamma^{\nu}\psi
\mathcal{O}(A)e^{i\int\ud^{4}z\mathcal{L}(\psi,A)(z)}
\end{eqnarray}
for quarks produced as color singlet and
\begin{eqnarray}
\Sigma^{\mu\nu}_{10}(x,y,\mathcal{O}(A))&=&
\frac{\big<T\{\bar{\psi}Y^{\dag}\gamma^{\mu}t^{a}Y\psi(x)\bar{\psi}Y^{\dag}\gamma^{\nu}Y\psi(y)
\bar{\psi}t^{a}\psi(0)\}
\big>}
{\int [\mathcal{D}\psi(z)][\mathcal{D}\bar{\psi}(z)][\mathcal{D} A^{\mu}(z)]
\bar{\psi}Y^{\dag}\gamma^{\mu}t^{a}Y\psi(x)\bar{\psi}Y^{\dag}\gamma^{\nu}Y\psi(y)
\bar{\psi}t^{a}\psi(0)
e^{i\int\ud^{4}z\mathcal{L}(\psi,A)(z)}}
\nonumber\\
&&
\int [\mathcal{D}\psi(z)][\mathcal{D}\bar{\psi}(z)][\mathcal{D} A^{\mu}(z)]\bar{\psi}\gamma^{\mu}Y^{\dag}t^{a}Y\psi(x)\bar{\psi}\gamma^{\nu}\psi
\mathcal{O}^{a}(A)e^{i\int\ud^{4}z\mathcal{L}(\psi,A)(z)}
\nonumber\\
&=&0
\end{eqnarray}
We see that, infrared divergences are topologically factorized in the color singlet case and vanish in the color octet case.

For the second diagram in Fig.\ref{infdiv}, infrared divergences cancel out in the summation over all possible undetected states $X$.

There are, of course, more complicated diagrams that contribute to fragmentation functions $D_{i\to \mu^{+}\mu^{-}(n,p_{H})}$ and $D_{Q\bar{Q}(\kappa)\to \mu^{+}\mu^{-}(n,p_{H})}$. For these diagrams, gluons that couple to nearly on shell intermediate particles which connect  two short distance(order $1/M$) subdiagrams do not produce pinch singular points in infrared region unless the relative velocities between these intermediate particles vanish(\cite{CN:1965,S:1993}). In the case that these relative velocities do vanish, we can repeat the analysis for the diagram Fig.\ref{selfenergy}. We consider the matrix-element:
\begin{eqnarray}
&&\big<X|T\{\mathcal{M}(Y_{n}\psi,Y_{n}\widetilde{A}Y_{n}^{\dag})(y)\bar{\psi}\gamma^{\mu}\psi(x)\mathcal{O}(A)\}|0\big>
\nonumber\\
&\propto&
\int_{(\psi(z),\bar{\psi}(z),\widetilde{A}^{\mu}(z),A^{\mu}(z))|_{z^{0}\to\infty }=X}
[\mathcal{D}\psi(z)][\mathcal{D}\bar{\psi}(z)][\mathcal{D}\widetilde{A}^{\mu}(z)][\mathcal{D} A^{\mu}(z)]
\nonumber\\
&&
\mathcal{M}(Y_{n}\psi,Y_{n}\widetilde{A}Y_{n}^{\dag})(y)
\bar{\psi}\gamma^{\mu}\psi(x)\mathcal{O}(A)
e^{i\int\ud^{4}z\mathcal{L}(\psi, \widetilde{A},A)(z)}
\end{eqnarray}
with the proportional coefficient independent of  the operator $\mathcal{O}(A)$, where $X$ represents any possible final states,  $\widetilde{A}$ represents the field corresponding to gluons with finite momenta, $Y_{n}$ is the Wilson line:
\begin{equation}
Y_{n}(x)=\mathcal{P}\exp(-ig \int_{0}^{\infty}\ud s n\cdot A(x^{\mu}+n^{\mu}))
\end{equation}
with $n^{\mu}$ the direction vector along the motion direction of the intermediate particles on the pinch surfaces,
$\mathcal{M}(Y_{n}\psi,Y_{n}\widetilde{A}Y_{n}^{\dag})$ represents effective operators that describe the interactions between infrared gluons and  intermediate particles that connect two short distance(order $1/M$) subdiagrams,    $\mathcal{O}(A)$ represents operators that describe  interactions between infrared gluons and the undetected particles $X$, the Lagrangian density $\mathcal{L}(\psi,A)(z)$ reads:
\begin{eqnarray}
\mathcal{L}(\psi,\widetilde{A},A)(z)&=&
\bar{\psi}(i\not\!\partial-m)\psi(z)
+\frac{1}{2g^{2}} tr ([\partial^{\mu}-ig\widetilde{A}^{\mu},\partial^{\nu}-ig\widetilde{A}^{\nu}])^{2}(z)
\nonumber\\
&&
+\frac{1}{2g^{2}} tr ([\partial^{\mu}-igA^{\mu},\partial^{\nu}-igA^{\nu}])^{2}(z)
\end{eqnarray}
. We do not consider effects of gauge fixing term and ghost fields for simplicity.  It is convenient for us to make the relative velocities between the intermediate particles slightly  deviate from 0 so that  Coulomb divergences caused by exchange of Coulomb gluons between these intermediate states, which do not affect the topological factorization, do not disturb us. We notice that:
\begin{eqnarray}
Y_{n}^{\dag}Y_{n}=1,\quad Y_{n}t^{a}Y_{n}=Y_{n}^{ab}t^{b}
\end{eqnarray}
.  Thus topologically unfactorized  infrared divergences are produced by the matrix element:
\begin{eqnarray}
\label{topunf}
S(0)&\equiv&\sum_{X}\big<0|\bar{T}\{\mathcal{M}^{\dag}(\ldots,Y_{n}^{ij}(0),\ldots,Y_{n}^{ab}(0),\ldots)\mathcal{O}^{\dag}(\ldots,Y_{l}(0),\ldots)\}|X\big>
\nonumber\\
&&
\big<X|T\{\mathcal{M}^{\dag}(\ldots,Y_{n}^{jk}(0),\ldots,Y_{n}^{bc}(0),\ldots)\mathcal{O}^{\dag}(\ldots,Y_{l}(0),\ldots)\}|0\big>
\nonumber\\
&=&\big<0|\bar{T}\{\mathcal{M}(\ldots,Y_{n}^{ij}(0),\ldots,Y_{n}^{ab}(0),\ldots)\mathcal{O}(\ldots,Y_{l}(0),\ldots)\}
\nonumber\\
&&
T\{\mathcal{M}(\ldots,Y_{n}^{jk}(0),\ldots,Y_{n}^{bc}(0),\ldots)\mathcal{O}(\ldots,Y_{l}(0),\ldots)\}|0\big>
\end{eqnarray}
. For the classical configurations $A^{\mu}(x)$, one can make use of unitarity to show that we can make the substitution
\begin{equation}
Y_{n}\to 1,\quad Y_{l}\to 1
\end{equation}
in (\ref{topunf}) without change the matrix element.

For quantum operators $A^{\mu}(x)$, we consider  Wilson lines:
\begin{equation}
W_{n}(0,s)\equiv\mathcal{P}\exp(-ig\int_{0}^{s}\ud\lambda n\cdot A(\lambda n^{\mu}))
\end{equation}
\begin{equation}
W_{l}(0,s)\equiv\mathcal{P}\exp(-ig\int_{0}^{s}\ud\lambda l\cdot A(\lambda l^{\mu}))
\end{equation}
. One can easily see that:
\begin{equation}
Y_{n}(0)=W_{n}(0,\infty),\quad Y_{l}(0)=W_{l}(0,\infty)
\end{equation}
. We consider the matrix element:
\begin{eqnarray}
S(0,s)&\equiv&\big<0|\bar{T}\{\mathcal{M}^{\dag}(\ldots,W_{n}^{ij}(0,s),\ldots,W_{n}^{ab}(0,s),\ldots)\mathcal{O}^{\dag}(\ldots,W_{l}(0,s),\ldots)\}
\nonumber\\
&&
T\{\mathcal{M}(\ldots,W_{n}^{jk}(0,s),\ldots,W_{n}^{bc}(0,s),\ldots)\mathcal{O}(\ldots,W_{l}(0,s),\ldots)\}|0\big>
\end{eqnarray}
We have:
\begin{eqnarray}
\frac{\ud}{\ud s}S(0,s)&=&ig
\big<0|\bar{T}\{\mathcal{M}^{\dag}(\ldots,W_{n}^{ij^{\prime}}(0,s),\ldots,W_{n}^{ab}(0,s),\ldots)\mathcal{O}^{\dag}(\ldots,W_{l}(0,s),\ldots)\}
n\cdot A_{j^{\prime}j}(sn^{\mu})
\nonumber\\
&&
T\{\mathcal{M}(\ldots,W_{n}^{jk}(0,s),\ldots,W_{n}^{bc}(0,s),\ldots)\mathcal{O}(\ldots,W_{l}(0,s),\ldots)\}|0\big>
\nonumber\\
&&-ig \big<0|\bar{T}\{\mathcal{M}^{\dag}(\ldots,W_{n}^{ij}(0,s),\ldots,W_{n}^{ab}(0,s),\ldots)\mathcal{O}^{\dag}(\ldots,W_{l}(0,s),\ldots)\}
\nonumber\\
&&
n\cdot A_{jj^{\prime}}(sn^{\mu})
T\{\mathcal{M}(\ldots,W_{n}^{j^{\prime}k}(0,s),\ldots,W_{n}^{bc}(0,s),\ldots)\mathcal{O}(\ldots,W_{l}(0,s),\ldots)\}|0\big>
\nonumber\\
&&+\ldots
\nonumber\\
&=&0
\end{eqnarray}
We see that:
\begin{equation}
S(0,s)=S(0,0)
\end{equation}
. Especially, we have:
\begin{equation}
S(0)=S(0,\infty)=S(0,0)=S(0)|_{A^{\mu}=0}
\end{equation}
.
We conclude that topologically unfactorized  infrared divergences cancel out.

We see that fragmentation functions $D_{i\to \mu^{+}\mu^{-}(n,p_{H})}$ and $D_{Q\bar{Q}(\kappa)\to \mu^{+}\mu^{-}(n,p_{H})}$ are free from topologically unfactorized infrared divergences  in the rest frame of the $\mu^{+}\mu^{-}$ pair. The detailed proof of such conclusion will be presented in other works.  We then have:
\begin{eqnarray}
D_{i\to \mu^{+}\mu^{-}(n,p_{H})}&=&D_{i\to \mu^{+}\mu^{-}(n,p_{H})}-D_{i\to \mu^{+}\mu^{-}(n,p_{H})}^{div}
\\
D_{Q\bar{Q}(\kappa)\to \mu^{+}\mu^{-}(n,p_{H})}&=&D_{Q\bar{Q}(\kappa)\to \mu^{+}\mu^{-}(n,p_{H})}-D_{Q\bar{Q}(\kappa)\to \mu^{+}\mu^{-}(n,p_{H})}^{div}
\end{eqnarray}
, where $D_{i\to \mu^{+}\mu^{-}(n,p_{H})}^{div}$ and $D_{Q\bar{Q}(\kappa)\to \mu^{+}\mu^{-}(n,p_{H})}^{div}$ represent topologically unfactorized  infrared divergences produced by various explicit diagrams in   $D_{i\to \mu^{+}\mu^{-}(n,p_{H})}$ and $D_{Q\bar{Q}(\kappa)\to \mu^{+}\mu^{-}(n,p_{H})}$. We can then make the decomposition:
\begin{eqnarray}
D_{i\to \mu^{+}\mu^{-}(n,p_{H})}&=&D_{i\to H}\otimes d_{H\to \mu^{+}\mu^{-}(n,p_{H})}
-(D_{i\to H}\otimes d_{H\to \mu^{+}\mu^{-}(n,p_{H})})^{div}
\nonumber\\
&&
+D_{i\to \mu^{+}\mu^{-}(n,p_{H})}-D_{i\to H}\otimes d_{H\to \mu^{+}\mu^{-}(n,p_{H})}
\nonumber\\
&&
-D_{i\to \mu^{+}\mu^{-}(n,p_{H})}^{div}+(D_{i\to H}\otimes d_{H\to \mu^{+}\mu^{-}(n,p_{H})})^{div}
\end{eqnarray}
\begin{eqnarray}
D_{Q\bar{Q}(\kappa)\to \mu^{+}\mu^{-}(n,p_{H})}&=&D_{Q\bar{Q}(\kappa)\to H}\otimes d_{H\to \mu^{+}\mu^{-}(n,p_{H})}
-(D_{Q\bar{Q}(\kappa)\to H}\otimes d_{H\to \mu^{+}\mu^{-}(n,p_{H})})^{div}
\nonumber\\
&&
+D_{Q\bar{Q}(\kappa)\to \mu^{+}\mu^{-}(n,p_{H})}-D_{Q\bar{Q}(\kappa)\to H}\otimes d_{H\to \mu^{+}\mu^{-}(n,p_{H})}
\nonumber\\
&&
-D_{Q\bar{Q}(\kappa)\to \mu^{+}\mu^{-}(n,p_{H})}^{div}+(D_{Q\bar{Q}(\kappa)\to H}\otimes d_{H\to \mu^{+}\mu^{-}(n,p_{H})})^{div}
\end{eqnarray}
, where $d_{H\to \mu^{+}\mu^{-}(n,p_{H})}$ represents the short distance(order $1/m$) evolution of the heavy quarkonium $H$ to the $\mu^{+}\mu^{-}$ pair, $D_{i\to H}$ and $D_{Q\bar{Q}(\kappa)\to H}$ represent the long distance evolution of the parton $i$ or the heavy quark pair $Q\bar{Q}$ to the hadron $H$ one concerned,  $(D_{i\to H}\otimes d_{H\to \mu^{+}\mu^{-}(n,p_{H})})^{div}$ and $(D_{Q\bar{Q}(\kappa)\to H}\otimes d_{H\to \mu^{+}\mu^{-}(n,p_{H})})^{div}$ represent the infrared divergences of the functions $D_{i\to H}$ and $D_{Q\bar{Q}(\kappa)\to H}$.
We have:
\begin{eqnarray}
&&D_{i\to H}\otimes d_{H\to \mu^{+}\mu^{-}(n,p_{H})}
-(D_{i\to H}\otimes d_{H\to \mu^{+}\mu^{-}(n,p_{H})})^{div}
\nonumber\\
&=&
(D_{i\to H}-D_{i\to H}^{div})\otimes d_{H\to \mu^{+}\mu^{-}(n,p_{H})}
\end{eqnarray}
\begin{eqnarray}
&&D_{Q\bar{Q}(\kappa)\to H}\otimes d_{H\to \mu^{+}\mu^{-}(n,p_{H})}
-(D_{Q\bar{Q}(\kappa)\to H}\otimes d_{H\to \mu^{+}\mu^{-}(n,p_{H})})^{div}
\nonumber\\
&=&
(D_{Q\bar{Q}(\kappa)\to H}-D_{QQ(\kappa)\to H}^{div})\otimes d_{H\to \mu^{+}\mu^{-}(n,p_{H})}
\end{eqnarray}
. Other terms in fragmentation functions $D_{i\to \mu^{+}\mu^{-}(n,p_{H})}$ and $D_{Q\bar{Q}(\kappa)\to \mu^{+}\mu^{-}(n,p_{H})}$ are suppress by the QED coupling constant $\alpha$ and the small decay width $\Gamma_{H}^{\mu^{+}\mu^{-}}$ of the heavy quarkonum $H$ to the $\mu^{+}\mu^{-}$ pair. We thus have:
\begin{eqnarray}
D_{i\to \mu^{+}\mu^{-}(n,p_{H})}&=&
(D_{i\to H}-D_{i\to H}^{div})\otimes d_{H\to \mu^{+}\mu^{-}(n,p_{H})}
\nonumber\\
&&
(1+O(\alpha)+O(\Gamma_{H}^{\mu^{+}\mu^{-}}))
\end{eqnarray}
\begin{eqnarray}
D_{Q\bar{Q}(\kappa)\to \mu^{+}\mu^{-}(n,p_{H})}&=&
(D_{Q\bar{Q}(\kappa)\to H}-D_{QQ(\kappa)\to H}^{div})\otimes d_{H\to \mu^{+}\mu^{-}(n,p_{H})}
\nonumber\\
&&
(1+O(\alpha)+O(\Gamma_{H}^{\mu^{+}\mu^{-}}))
\end{eqnarray}
.

It is interesting to mention that intermediate sates which connect two short distance(order $1/M$) subdiagrams do not produce infrared divergence terms unless relative velocities of these states vanish. Thus such infrared divergences occur only when the momentum of the final $\mu^{+}\mu^{-}$ pair $p_{H}$ take some special values. Such infrared divergences vanish in the integral of $p_{H}$ once infrared divergences are no worse than logarithms. For the cancellation of topologically unfactorized divergences, however the integral over $p_{H}$ is not necessary.

\section{Conclusion}
\label{conclusion}

We present a scheme to cancel out topologically unfactorized infrared divergences in inclusive productions of heavy quarkonia.  In \cite{Bodwin:1994jh}, it is proposed that such divergences  do cancel out according to the KLN(\cite{KLN:1962,KLN:1964}) cancellation once the summation over the undetected particles is made. The explicit calculations at NNLO in \cite{NQS:2005,Nayak:2006fm} display that, however, the summation is not inclusive enough to cancel out topologically unfactorized infrared divergences.

In \cite{NQS:2005,Nayak:2006fm}, the final states are chosen as color singlet heavy quark pair plus any other states with the final heavy quark and heavy antiquark both on shell. We notice that, the color singlet heavy quark pair is not invariant under the evolution of infrared QCD interactions. Especially, color states of the heavy quark pair may change to others under such evolution. Thus the KLN cancellation, for which the summation over all states that arise from such evolution is necessary, does not work simply. In fact, practical heavy quarkonia is not the color singlet heavy quark pair. It seems plausible to define Heavy quarkonia as resonance states which are invariant under the evolution of infrared QCD interactions as we do for the $J/\psi$ particle. Thus summation over higher Fock states is necessary. Such summation, however, do not make topologically unfactorized infrared divergences disappear as shown in our calculations.

It is interesting to point it out that the states $HX$ do not form the invariant subspace of the evolution of infrared QCD interactions even if the detected particle $H$ is itself invariant under such evolution. Exchanges  of soft gluons between $H$ and $X$ may change the state $H$, for example, cause the transition between heavy quarkonia. Such transition may exist even if all soft gluons are infrared! We notice that, Heavy quarkonia are reconstructed by their decay products in practical experiments. Final decay products like $\mu^{+}\mu^{-}$,  the invariant mass of which are require to be near  the mass of the detected heavy quarkonium $H$, can be produced by the decay of $H$ and other possible states.  Thus the practical process is indeed inclusive about the states arising from the evolution of infrared QCD interactions. We show that topologically unfactorized infrared divergences do cancel out in fragmentation functions $D_{i\to \mu^{+}\mu^{-}(n,p_{H})}$ and $D_{Q\bar{Q}(\kappa)\to \mu^{+}\mu^{-}(n,p_{H})}$ in this paper!

It seems more reasonable to consider the NRQCD factorization for fragmentation functions $D_{i\to \mu^{+}\mu^{-}(n,p_{H})}$ and $D_{Q\bar{Q}(\kappa)\to \mu^{+}\mu^{-}(n,p_{H})}$ as they are free from topologically unfactorized infrared divergences.
If the NRQCD factorization theorem holds for these fragmentation functions, then we have:
\begin{eqnarray}
D_{i\to \mu^{+}\mu^{-}(n,p_{H})}&=&
\sum_{n}(d_{i\to Q\bar{Q}(n)}-d_{i\to Q\bar{Q}(n)}^{div})(\big<\mathcal{O}^{H}(n)\big>-\big<\mathcal{O}^{H}(n)\big>^{div})
\nonumber\\
&&
\otimes (d_{H\to \mu^{+}\mu^{-}(n,p_{H})}-d_{H\to \mu^{+}\mu^{-}(n,p_{H})}^{div})
\nonumber\\
&&
(1+O(\alpha)+O(\Gamma_{H}^{\mu^{+}\mu^{-}}))
\end{eqnarray}
\begin{eqnarray}
D_{Q\bar{Q}(\kappa)\to \mu^{+}\mu^{-}(n,p_{H})}&=&
\sum_{n}(d_{Q\bar{Q}(\kappa)\to Q\bar{Q}(n)}-d_{Q\bar{Q}(\kappa)\to Q\bar{Q}(n)}^{div})(\big<\mathcal{O}^{H}(n)\big>-\big<\mathcal{O}^{H}(n)\big>^{div})
\nonumber\\
&&
\otimes (d_{H\to \mu^{+}\mu^{-}(n,p_{H})}-d_{H\to \mu^{+}\mu^{-}(n,p_{H})}^{div})
\nonumber\\
&&
(1+O(\alpha)+O(\Gamma_{H}^{\mu^{+}\mu^{-}}))
\end{eqnarray}
.
We do not require that  matrix elements $\big<\mathcal{O}^{H}(n)\big>$ absorb all infrared divergences in fragmentation functions $D_{i\to H}$ and $D_{Q\bar{Q}(\kappa)\to H}$. Generally speaking, parts of infrared infrared divergences in functions $D_{i\to H}$ and $D_{Q\bar{Q}(\kappa)\to H}$ are cancelled by infrared divergences of other terms in fragmentation functions $D_{i\to \mu^{+}\mu^{-}(n,p_{H})}$ and $D_{Q\bar{Q}(\kappa)\to \mu^{+}\mu^{-}(n,p_{H})}$. We thus obtain a NRQCD factorization theorem for the inclusive production of the heavy quarkoum $H$ once the  NRQCD factorization for fragmentation functions $D_{i\to \mu^{+}\mu^{-}(n,p_{H})}$ and $D_{Q\bar{Q}(\kappa)\to \mu^{+}\mu^{-}(n,p_{H})}$ hold.

\section*{Acknowledgements}

The author thanks Y. Q. Chen for for helpful discussions and important suggestions about the work.

\bibliography{inclusiveproduction}

\end{document}